\documentclass[longauth]{aa}

\usepackage{graphicx}
\usepackage{xcolor}
\usepackage{subcaption}

\usepackage{txfonts}
\usepackage[flushleft]{threeparttable}
\usepackage{hyperref}
\usepackage{amsmath} 
\usepackage{bm}
\DeclareUnicodeCharacter{2212}{-}

\begin{document}

   \title{Spatially Resolved Star Formation relations in local LIRGs along the complete merger sequence}

   \author{M. S\'anchez-Garc\'ia
          \inst{1,2}
          \and
          T. D\'iaz-Santos\inst{1,2,3}
           \and
           L. Barcos-Muñoz\inst{4,5}
           \and
           A. S. Evans\inst{4,5}
           \and
           Y. Song\inst{6,7}
           \and
           M. Pereira-Santaella\inst{8}
           \and
           S. Garc\'ia-Burillo\inst{9}
           \and
           S. T. Linden\inst{10}
           \and
           C. Ricci\inst{11,12}
            \and
           L. Lenkic\inst{13}
           \and
           A. Zanella\inst{14}
           \and
           L. Armus\inst{13}
           \and
           C. Eibensteiner\inst{4}\thanks{Jansky Fellow of the National Radio Astronomy Observatory}
           \and
           Y.-H. Teng\inst{15}
           \and
           A. Saravia\inst{5}
           \and
           V. A. Buiten\inst{16}
           \and
           G. C. Privon\inst{4,5,17}
            \and
            N. Torres-Albà\inst{5}
             \and
            T. Saito\inst{18}
             \and
        K.~L.~Larson\inst{19}
           \and
            M. Bianchin\inst{20}
             \and
A. M. Medling\inst{21}
              \and
             T. Lai\inst{22}
             \and
           G. P. Donnelly\inst{23}
           \and
           V. Charmandaris\inst{1,2,3}
            \and
            T. Bohn\inst{24}
              \and
           C. M. Lofaro\inst{1,2}
           \and
           G. Meza\inst{25}
           }

   \institute{
    Institute of Astrophysics, Foundation for Research and Technology-Hellas (FORTH), Heraklion, 70013, Greece \\\email{msanchez@ia.forth.gr}
    \and  
    Department of Physics, University of Crete, Heraklion, 71003, Greece 
    \and
    School of Sciences, European University Cyprus, Diogenes Street, Engomi, 1516, Nicosia, Cyprus
    \and
    National Radio Astronomy Observatory, 520 Edgemont Road, Charlottesville, VA 22903, USA 
    \and
    Department of Astronomy, University of Virginia, 530 McCormick Road, Charlottesville, VA 22903, USA
   \and
    European Southern Observatory, Alonso de C\'ordova, 3107, Vitacura, Santiago 763-0355, Chile
    \and
    Joint ALMA Observatory, Alonso de C\'ordova, 3107, Vitacura, Santiago 763-0355, Chile
    \and
 Instituto de F\'isica Fundamental, CSIC, Calle Serrano 123, 28006 Madrid, Spain 
    \and
    Observatorio Astron\'omico Nacional (OAN-IGN)-Observatorio de Madrid, Alfonso XII, 3, 28014 Madrid, Spain
    \and
    Steward Observatory, University of Arizona, 933 N Cherry Avenue, Tucson, AZ 85721, USA
    \and
    Instituto de Estudios Astrofísicos, Facultad de Ingeniería y Ciencias, Universidad Diego Portales, Av. Ejército Libertador 441, Santiago, Chile
    \and
    Department of Astronomy, University of Geneva, ch. d'Ecogia 16, 1290, Versoix, Switzerland
    \and
    IPAC, California Institute of Technology, 1200 E. California Blvd., Pasadena, CA 91125, USA
   \and
    INAF – Osservatorio di Astrofisica e Scienza dello Spazio di Bologna, Via Gobetti 93/3, 40129, Bologna, Italy
    \and
    Department of Astronomy, University of Maryland, 4296 Stadium Drive, College Park, MD 20742, USA
    \and
   Leiden Observatory, Leiden University, PO Box 9513, 2300 RA Leiden, The Netherlands
    \and
    Department of Astronomy, University of Florida, 1772 Stadium Road, Gainesville, FL 32611, USA
    \and
    Faculty of Global Interdisciplinary Science and Innovation, Shizuoka University, 836 Ohya, Suruga-ku, Shizuoka 422-8529, Japan
    \and
    AURA for the European Space Agency (ESA), Space Telescope Science Institute, 3700 San Martin Drive, Baltimore, MD 21218, USA
    \and
    Department of Physics and Astronomy, University of California, 4129 Frederick Reines Hall, Irvine, CA 92697, USA
    \and
    Department of Physics \& Astronomy and Ritter Astrophysical Research Center, University of Toledo, Toledo, OH 43606,USA
    \and
    zz
    \and
    Ritter Astrophysical Research Center, University of Toledo, Toledo, OH 43606, USA
    \and 
    Hiroshima Astrophysical Science Center, Hiroshima University, 1-3-1 Kagamiyama, Higashi-Hiroshima, Hiroshima 739-8526, Japan
    \and
    Universidad Nacional Autónoma de Honduras, Ciudad Universitaria, Tegucigalpa, Honduras
    }

  \abstract
   {We investigate the properties of the interstellar medium (ISM) at giant molecular cloud (GMC) scales ($\sim$ 100 pc) in a sample of 27 nearby luminous infrared galaxies (LIRGs) spanning all interacting stages along the merger sequence; i.e., from isolated systems to late-stage mergers. In particular, we study the relations between star-formation (SF) and molecular gas surface density as a function of the interaction stage by (1) defining beam-sized (unresolved, line-of-sight) regions and (2) identifying actual gas clumps and physical structures within the galaxies. In total, we identify more than 4000 beam-sized CO-emitting regions defined on scales of $\sim$100 pc and more than 1000 molecular gas clumps in the sample. To map the distribution of molecular gas we use ALMA to observe the J = 2--1 CO transition, and to map the distribution of star formation we use HST observations of the Pa$\alpha$ or Pa$\beta$ hydrogen recombination lines.  We derive spatially resolved Kennicutt–Schmidt (KS) relations for each LIRG in the sample.
  When using beam-sized regions, we find that 67\% of galaxies follow a single relation between $\Sigma_{SFR}$ and $\Sigma_{H2}$. However, in the remaining galaxies, the relation splits into two branches -- one is characterised by higher $\Sigma_{SFR}$ and $\Sigma_{H2}$, while the other exhibits lower values -- indicating the presence of a duality in this relation. In contrast, when using physical gas clumps, the duality disappears and all galaxies show a single trend.
  These results provide two complementary perspectives when studying the star formation process: one that maximises the number statistics (beam-sized regions), and another that focuses on actual structures associated to gas clumps in which the measured sizes have a physical meaning. We also study other ISM/clump properties as a function of the merger stage of the LIRG systems. 
  We find that isolated galaxies and systems in early stages of interaction exhibit lower amounts of gas and star formation. As the merger progresses, however, the amount of gas in the central kiloparsecs of the galaxy undergoing the merger increases, along with the SFR, and the slope of the KS relation becomes steeper, indicating an increase in the SF efficiency of the molecular gas clumps. Clumps in late-stage mergers are predominantly located at small distances from the nucleus, confirming that most of the activity is concentrated in the central regions. Most interestingly, the relation between the star formation efficiency and the boundedness parameter (which measures the effects of gravity against velocity dispersion) evolves from being roughly flat in the early stages of the merger to becoming positive in the final phases, indicating that clump self-gravity only starts to regulate the star formation process between the early- and mid-merger stages. 
  }


   \maketitle

\section{Introduction}

Star formation (SF) takes place primarily within structures of
molecular gas, called Giant Molecular Clouds (GMCs).
Their formation, evolution, and lifetimes are shaped by a combination of internal processes -- such as self-gravity and stellar feedback -- and external environmental factors like galactic dynamics and pressure (e.g., \citealt{Bournaud2015}; \citealt{wilson2019}; \citealt{SchinnererLeroy2024}; \citealt{Meidt2025}).
Stellar feedback, in the form of stellar winds, supernova explosions, and radiation pressure from massive stars, injects energy and momentum into the interstellar medium (ISM), driving turbulence and potentially disrupting molecular clouds \citep[e.g.][]{Renaud2019, chevance2022, bonne2023}.  
In some cases, feedback-induced compression or dynamical shear may promote cloud collapse and subsequent SF \citep[e.g.][]{orr2018, chevance2020}.
It remains under debate whether SF is primarily governed by internal properties such as cloud self-gravity, or by external environmental conditions -- such as galactic shear, pressure, or tidal forces -- that regulate the efficiency and spatial distribution of SF \citep[e.g.][]{Kruijssen2014, corbelli2017, reyraposo2017, liu2022, Choi2023, sun2023, cenci2024}.

The low-J CO transitions have been used for decades as the primary tracers of molecular gas in galaxies. Early CO observations, limited by low spatial resolution, could not resolve individual GMCs, detecting instead kiloparsec-scale complexes of emission. As a result, early studies probed star formation at global or averaged scales, leading to the empirical Kennicutt–Schmidt (KS) relation \citep{Schmidt59, Kennicutt1998bis}, which links star formation and gas surface densities.
Despite its success as an empirical law, the physical mechanisms that govern star formation remain unknown. Understanding these mechanisms requires resolving GMCs and linking their properties to recent star formation across a variety of galactic environments.

Observations of molecular clouds in our Galaxy and in a number of nearby galaxies have identified various empirical trends manifesting such cloud–environment correlations.
Within a galaxy, molecular clouds located closer to the galaxy centre appear denser, more massive, and more turbulent (e.g. \citealt{Oka2001, Colombo2014, Freeman2017, Hirota2018, Brunetti2021}, also see \citealt{Heyer2015}). Similar trends have been found in galaxy-scale numerical simulations \citep[e.g.][]{Jeffreson2020}. Recent observational works also report that more massive and actively star-forming galaxies tend to host clouds with typically larger sizes, masses, surface densities, and velocity dispersions (\citealt{Leroy2015, Leroy2016, Schruba2019, Sun2020}, see also \citealt{Bolatto2008, Fukui2010}). However, galaxy mergers, which represent a crucial phase in galaxy evolution, have been scarcely studied, particularly in relation to GMCs \citep{Brunetti2021, Brunetti2022, Brunetti2024, He2024}. These events play a central role in shaping the growth, star formation history, and overall evolution of galaxies, influencing their dynamics and structure throughout the history of the universe.

Previous studies of cloud-scale star formation were typically limited to a small number of galaxies or sub-galactic regions, restricting the diversity of physical environments probed.
While these studies provided unique insights for specific targets, only systematic large samples of galaxies can cover a wide range of host galaxy properties, produce representative population statistics, and establish meaningful connections with galaxy evolution models. The PHANGS survey marked a turning point by enabling such studies in star-forming main-sequence galaxies \citep{Sun2018, Sun2022, Utomo2018, Leroy2021, Pessa2021, Kim2022}.

Multi-wavelength observations have shown that local luminous and ultraluminous infrared galaxies (LIRGs, with L$_{IR}$=10$^{11-12}$L$_{\odot}$, and ULIRGs with  L$_{IR}$>10$^{12}$L$_{\odot}$) are a mixture of single disc galaxies, interacting systems, and advanced mergers, exhibiting enhanced SFRs and active galactic nucleus (AGN) activity compared to less luminous and non-interacting galaxies \citep[c.f.][]{sanders96, Stierwalt2013, Larson2016}. 
These extreme environments are particularly relevant for testing the Kennicutt–Schmidt relation, as both low- and high-resolution studies suggest that U/LIRGs deviate from the typical behaviour of normal star-forming galaxies \citep[e.g.][]{daddi2010, genzel2010, Burillo2012, SG2022, sg2022a,saravia2025}.
As merger-driven systems, they also offer unique insights into the interplay between star formation, dynamics, and galactic evolution \citep[e.g.][]{Hopkins2006}. 

Given their high dust content, investigating star formation in LIRGs requires infrared observations that mitigate obscuration while preserving spatial detail. Hydrogen recombination lines in the near-infrared -- such as Pa$\alpha$, Pa$\beta$, or Br$\gamma$ -- are well suited for this purpose, as their emission directly traces ionising photons from massive O/B stars. In contrast, optical lines like H$\alpha$ and H$\beta$ are severely attenuated in LIRG nuclei, where visual extinctions can exceed A$_{V}$ > 10 \citep{Armus1989, garcia-marin2009,piqueras2013, Stierwalt2013}. 
The lower extinction at near-IR wavelengths (e.g. 1.6 $\mu$m $\sim$ 0.2 $\times$ A$_{V}$) allows for a more complete census of obscured star formation.

In this paper, we present high-resolution (48–112 pc) observations obtained by the Atacama Large Millimetre Array (ALMA) in the 2--1 transition of CO to characterise the properties of GMCs and the local ISM in a sample of nearby LIRGs, which span the entire merger sequence. The star formation associated with the GMCs is estimated from the Hubble Space Telescope (HST) near-infrared (NIR) recombination line images. 
 With these data, we study SF relations and the effects of the merger environment on the scales of conventionally defined massive GMCs or giant HII regions
\citep[e.g.][]{Miville2017}.

\section{Sample and observations} 
\subsection{Sample selection}\label{sample}
We present sub-kpc CO(2--1) observations obtained by ALMA of a representative sample of 27 local LIRGs, which has archival NIR observations from HST. 
Our sample is drawn from the Great Observatories All-sky LIRG Survey \citep[GOALS;][]{Armus2009}. GOALS is a complete galaxy sample that comprises the 201 LIRG systems 
(L$_{IR}\geq$10$^{11}$L$_{\odot}$ and f$_{60\mu m}$> 5.24 Jy) in the IRAS Revised Bright Galaxy Sample \citep{Sanders2003} and is aimed at measuring the physical properties of local LIRGs across the electromagnetic spectrum using a broad suite
of ground- and space-based observatories.

\begin{figure}[ht!]
   \centering
    \includegraphics[width=.95\linewidth]{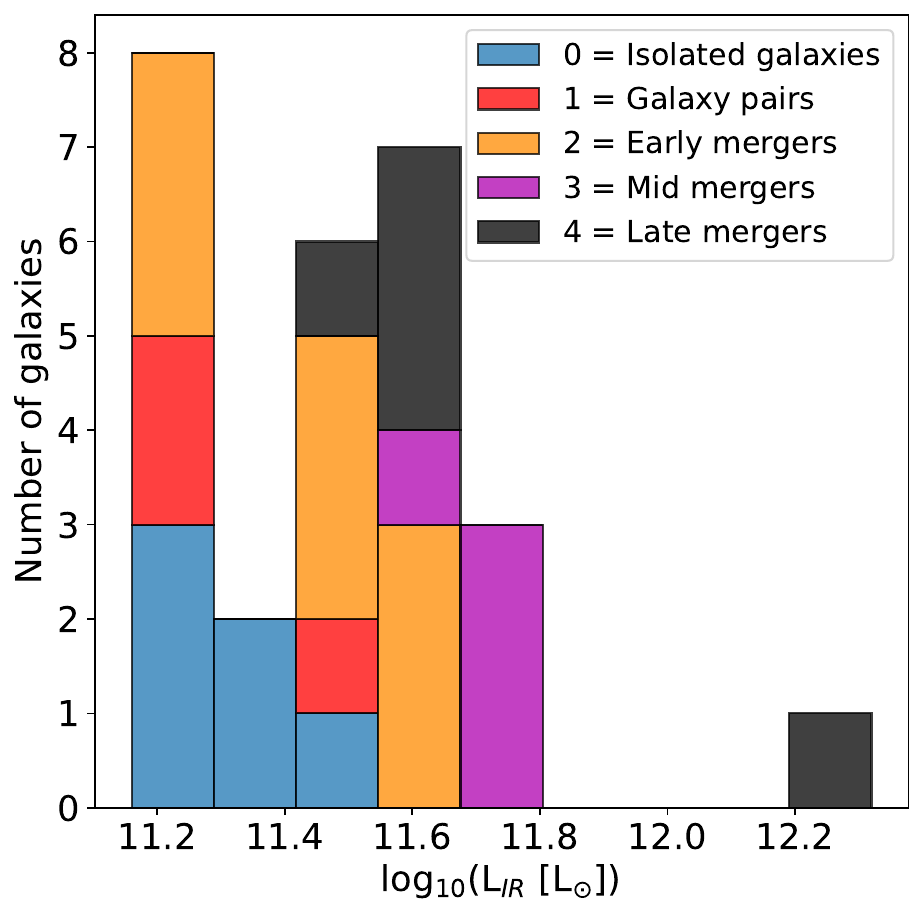}
 \caption{
 Stacked histogram showing the distribution of infrared luminosity (L$_{IR}$) across the different merger stages of the individual galaxies in the sample. The early stages of the merger sequence tend to have lower infrared luminosities, while galaxies in more advanced merger stages predominantly show higher infrared luminosities. 
 }
    \label{histogram:lir}
\end{figure}

The merger stage classification of our sample is based on \cite{Stierwalt2013}, except for three objects (IRAS08355--4944, IRASF13497+0220, IRAS15206--6256), which we have reclassified. 
The reason for the modification is that the Spitzer classification for these galaxies differs from that based on HST data. While the morphology of some systems in the Spitzer images led to their classification as late-stage mergers, the higher resolution HST data clearly show that the interacting galaxies are in an early merger stage, with minimally disturbed morphologies and no common envelope of stars (the key signature of late-stage mergers). Our sample spans a range of log $L_{IR}$ from 11.16 to 11.73 $L_{\odot}$, except for one galaxy classified as a ULIRG, with 12.32 $L_{\odot}$ (see Fig. \ref{histogram:lir} and Column 8 in Table \ref{tab:sample}). Five objects are classified as AGN and/or show evidence of AGN activity. In Table \ref{tab:sample} we present the main properties of the individual galaxies in the sample.

\begin{table*}[ht!]
    \centering	
	\caption{General properties of the sample of Local LIRGs}

    \resizebox{\linewidth}{!}{
	\begin{tabular}{ l l c c c c c c c c c c}
	\hline
	\hline
	\multicolumn{2}{c}{Object}& & $\alpha$ & $\delta$ & $z$ & D$_{L}$ & \textit{i} &  log~$L_{IR}$ & Spectral & Ref. & Stage  \\
		
	\cline{1-2} \cline{4-5}
	IRAS & Galaxy Name  & & J2000.0 & J2000.0 &   & & &  & Class &  &  \\ 
	
	&  & & [h m s] & [$^{\circ}$  $\arcmin$ $\arcsec$] & & [Mpc] & [$^{\circ}$] & [L$_{\odot}$] & & & \\ 

	(1) & (2) & & (3) & (4) & (5) & (6) & (7) & (8) & (9) & (10) & (11)  \\
	\hline
		
	F00163--1039 N & MCG~--02--01--052 & & 00 18 49.85 & -10 21 34.00 & 0.0272 & 117.5 & 27 $\pm$ 1&  11.48 &  H & 1,2 & 2 \\

	F00163--1039 S &  MCG~--02--01--051 &  &00 18 50.90 & -10 22 36.7 & 0.0272 & 117.5 &64 $\pm$ 5 &  11.48 &  H & 1, 2 &  2\\
    
    F04315--0840  & NGC~1614 & &04 33 59.95 &  -08 34 46.6 & 0.0159 & 69.7 & 48 $\pm$ 2  & 11.65 & C & 3 & 4\\
    
    F07160--6215  & NGC~2369 & &07 16 37.73 & -62 20 36.4 &  0.0108 & 49.7 & 66 $\pm$ 6  &11.16  & C  & 3,4 & 0 \\
    
    08355--4944 & &  &08 37 01.87 & -49 54 30.0 & 0.0259  &118.0 & 39 $\pm$ 5   &11.62 & - & 5 & 3\\
    
    F10015--0614 & NGC~3110 & &10 04 02.11 & -06 28 29.5 &  0.0169  & 79.5 & 57 $\pm$ 3 & 11.42 & H &  3, 5, 6 & 1 \\
  
    F10257--4339  & NGC~3256 & & 10 27 51.30 &  -43 54 14.0 & 0.0094 & 45.7 & - &11.64  & H & 3,4, 5,6 & 4 \\

    F11506--3851 & ESO~320-G030 & & 11 53 11.73 & -39 07 49.0 & 0.0108 & 41.2 & 56 $\pm$ 4  &11.17 & H & 3,4, 5,6 & 0 \\
        
    F12596--1529 W& MCG~-02-33-098 W & & 13 02 19.66 & -15 46 04.2 & 0.0159 & 78.7 & 54$\pm$ 6 & 11.17 &  H & 1,5,6 & 2 \\
    
    F12596--1529 E& MCG~-02-33-098 E& & 13 02 20.38 & -15 45 59.6 & 0.0159 & 78.7 & 39 $\pm$ 1 & 11.17  &  H & 1,3, 5,6 &2\\
    
    13120--5453 & & &13 15 06.37 & -55 09 22.5 &  0.0308 &  144.0 & 35 $\pm$ 6  &  12.32 & Sy2 & 6 &4\\
    
    F13182+3424 & UGC~08387 & &13 20 35.37 & +34 08 22.2 & 0.0233 & 110.0 &48 $\pm$ 4 &  11.73 & L & 2 &3 \\
    
    F13229--2934 & NGC~5135 && 13 25 44.02 & -29 50 00.4 & 0.0137 & 60.9 & 53 $\pm$ 9 &11.30 & Sy2 & 1, 3,6,7 &0 \\
    
    F13370+0105 E& NGC~5258 && 13 39 57.72  & +00 49 53.0 & 0.0226 & 108.5 & 67 $\pm$ 3  & 11.22 & H/L & 8  &2 \\
    
    F13370+0105 W & NGC~5257 && 13 39 52.95  & +00 50 25.9 & 0.0226 & 108.5 & 45 $\pm$ 14  & 11.50  & H &  8 &2 \\
    
    F13497+0220 S& NGC~5331 && 13 52 16.21 & +02 06 05.1 & 0.0330 & 155.0 & 58 $\pm$ 2  & 11.66  &-  &9& 2\\
    
    F13497+0220 N& NGC~5331& & 13 52 16.43 & +02 06 30.9 & 0.0330 & 155.0 & 66 $\pm$ 3 & 11.66  & H & 9&2 \\
    
    F14544--4255 W & IC~4518 W& &  14 57 41.22 & -43 07 55.8 &  0.0159 &  74.6 & 50 $\pm$ 4 &  11.24  & Sy2  & 3,7 &1 \\
    
    F14544--4255 E & IC~4518 E& & 14 57 45.33 & -43 07 57.0 & 0.0159 &  71.2 &  75 $\pm$  2 &  11.24  & H & 6& 1 \\
    
    15206--6256 S& ESO~099-G004&& 15 24 57.98 & -63 07 29.4 & 0.0293 & 137.0 & 58 $\pm$ 2 & 11.74  & & &3\\
    
    15206--6256 N& ESO~099-G004& &15 24 58.98 & -63 07 29.4 & 0.0293 & 137.0 & 59 $\pm$ 1  & 11.74  & & &3 \\
    
    F16164--0746 & && 16 19 11.75 & -07 54 03.0 &  0.0272 & 128.0 & 47 $\pm$ 3 &  11.62 & L & 6 & 4 \\
   
    F17138--1017 &  && 17 16 35.68 &  -10 20 40.5 &  0.0173 & 76.7 & 50 $\pm$ 1  &11.49  & H  & 8 & 4 \\

    F18341--5732 & IC~4734 & &18 38 25.75 & -57 29 25.4 & 0.0157 & 73.4& 58 $\pm$ 10  &11.35  & H & 3,6  & 0 \\

    F21453--3511 & NGC~7130 & &21 48 19.54 & -34 57 04.7 & 0.0162 &  67.6 &  50 $\pm$ 9 &  11.42   & Sy2 & 6,7 &0\\

    F22132--3705 & IC~5179 & &22 16 09.13 & -36 50 37.2 & 0.0114 & 51.4 & 62 $\pm$ 5  &11.24  & H & 3,6 &0 \\

    F23007+0836 & NGC~7469& & 23 03 15.64 & +08 52 25.5 & 0.0163 & 70.8 &  39 $\pm$ 5 &11.65 & Sy1 & 3,6 &2  \\ [1ex]
    \hline
	 		
	\end{tabular}}
\tablefoot{Col. (1): IRAS system name, where an “F” prefix indicates the Faint Source Catalogue and no prefix indicates the Point Source Catalogue. Col. (2): Galaxy name. Col. (3) and (4): right ascension (hours, minutes, seconds) and declination (degrees, arcminutes, arcseconds) from \cite{Stierwalt2013}. Col. (5): Redshift \citep{Stierwalt2013}. Col. (6): Luminosity distance from \cite{Armus2009}. Col. (7): Inclination. Col. (8): Infrared luminosity (L$_{IR}$(8--1000 $\mu$m)) calculated from the IRAS flux densities f$_{12}$, f$_{25}$, f$_{60}$, and f$_{100}$ \citep{Stierwalt2013}. Col. (9): Nuclear activity optical spectral class (H = HII region-like, L = LINER, Sy2 = Seyfert 2, Sy1 = Seyfert 1 and C = Composite). Col. (10): References of the nuclear classification: 1: \cite{Kewley2001}; 2: \cite{Veilleux1995}; 3: \cite{Alonso-Herrero2012};  4:\cite{Pereira2011}; 5: \cite{RZ2011};  6:\cite{Pereira2010b};   7: \cite{Corbett2003}; 8:\cite{Yuan2010}; 
9: \cite{iono2005}. 
Col. (11): Merger stage based on visual classification: 0 = isolated galaxies, 1 = pairs of galaxies; 2 = early-stage merger, 3 = mid-stage merger, 4 = late-stage merger. }
	
\label{tab:sample}
\end{table*}

\subsection{CO(2--1) ALMA data}\label{co:observations}
We used ALMA Band 6 observations of the CO(2--1) emission line from ALMA proposal 2017.1.00395.S (PI: T. D\'iaz-Santos) and completed the sample with the data described in \cite{SG2022}. The total sample covers the complete merger sequence of galaxies, where most of the galaxies from \cite{SG2022} are isolated galaxies and early-stage mergers. 
The data were calibrated using the standard ALMA data reduction software,  CASA\footnote{\href{http://casa.nrao.edu/}{http://casa.nrao.edu/}}
\citep{McMullin2007}. We subtracted the continuum emission in the \textit{uv} plane using an order 0 baseline. For the cleaning we used the Briggs weighting with a robustness parameter of 0.5 \citep{Briggs1995}, providing a spatial resolution of 48–112 pc. From the 23 LIRGs observed (a total of 27 individual galaxies in the sample when all the galaxies in pairs are counted), 13 were observed using two configurations, compact and extended, and 10 using only the extended configuration. The maximum recoverable scale (MRS) for the compact plus extended configuration data ranges between $\sim$6$\arcsec$ and $\sim$11$\arcsec$ ($\sim$1.4 and 3.5 kpc). In the case of the extended-only configuration observations, the MRS is $\sim$3$\arcsec$ ($\sim$1.1 kpc).  In this paper we study spatial scales of $\sim$100 pc, which are 10 times smaller than the MRS, so we expect the missing flux due to the absence of short spacing to be low at these scales. In addition, for a few objects of this sample with single-dish CO(2--1) observations, the integrated ALMA and single-dish fluxes agree within 15\% \citep{pereira2016b, pereira2016}. The final data cubes have channels of 7.8 MHz ($\sim$10 km s$^{-1}$) for the sample. The field of view (FoV) of the ALMA single pointing data has a diameter of $\sim$25$\arcsec$ ($\sim$5--16 kpc). The mosaics have a diameter between $\sim$38$\arcsec$ and $\sim$64$\arcsec$ ($\sim$11 and 34 kpc). We applied the primary beam correction to the data cubes. Further details on the observations for each galaxy are listed in Table \ref{tab:almainfo}.

A common spatial scale of $\sim$90 pc was defined in order to have a homogeneous data set. 
This physical scale was chosen to preserve the original resolution of the nearest objects as much as possible, with minimum degrading, while accommodating only a few most distant sources with a slightly coarser resolution than the common spatial scale.
We convolved to 90 pc the data cubes of 30\% of the sample with spatial resolutions higher than 80 pc. 
There are 5 objects (three sources where two of them are galaxy pairs), representing 19\% of the sample, with an original resolution of up to $\sim$110 pc.
For these remaining objects, we directly used the cleaned data cubes at their original spatial resolution. For consistency and clarity, we therefore refer throughout the paper to a characteristic resolution of $\sim$100 pc. 

\begin{figure*}[ht]
   \centering
    \includegraphics[width=0.9\linewidth]{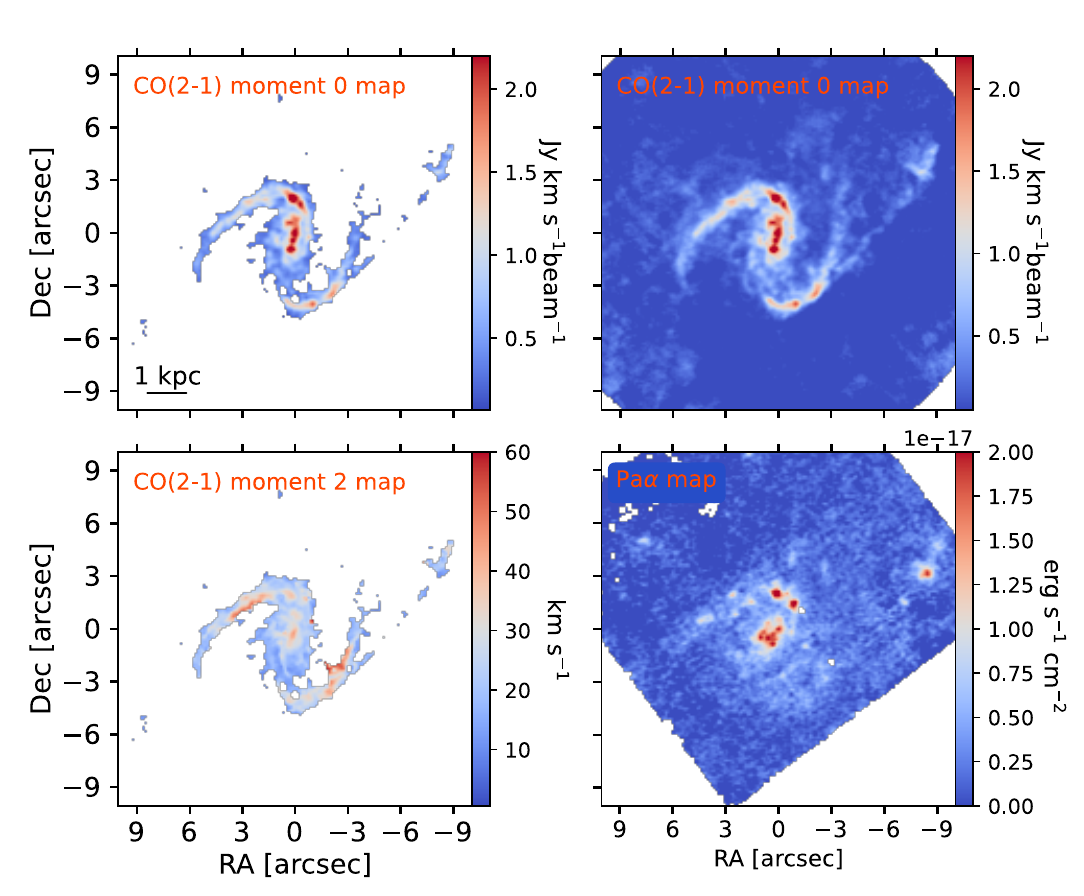}
 \caption{ALMA CO(2--1) integrated intensity (moment 0) maps, with pixel masking applied to the data cube (top left panel) and without it (top right panel), CO(2--1) velocity dispersion (moment 2) map (bottom left panel) and the HST Pa$\alpha$ image (bottom right panel) of the galaxy NGC\,3110 at $\sim$100 pc scale. At this scale, significant differences are observed between the CO(2--1) moment 0 and Pa$\alpha$ maps, revealing emission structures distributed throughout the galaxy.
 }
    \label{fig:maps}
\end{figure*}

We obtained the CO(2--1) moment 0 and 2 maps (see the top and bottom--left panels of Figure \ref{fig:maps}) using two different methodologies: i) masking pixels in each channel map with fluxes < 3$\times$RMS$_{CO}$, where RMS$_{CO}$ is the background noise (for more details, see \cite{SG2022}; the code is available on GitHub\footnote{\href{https://github.com/itsmariasg/moment_maps}{https://github.com/itsmariasg/moment\_maps}}); and ii) without applying any masking or clipping to the data. We use the first method to follow the methodology of \cite{SG2022} and extend the study to a larger sample of galaxies. With this method, we obtain the brightest emission while being restrictive with the noise in the data, whereas we use the second method to include the fainter regions of the clumps.

\begin{table*}[ht!]
    \centering
	\caption{ALMA CO(2--1) observations of the sample.}
	\begin{threeparttable}
	\begin{tabular}{ l c c c c c c c c c}
	\hline
	\hline
	Object & $\theta_{maj}\times\theta_{min}$ & $\theta_{m}$ &  P.A. & Sensitivity & Mosaics & MRS  & FoV & Project & HST image\\
		
	IRAS Name & [$\arcsec$]~~~~~~[$\arcsec$] & [$\arcsec$, pc] & [$^{\circ}$] & [mJy beam$^{-1}$] & & [$\arcsec$] & [$\arcsec$] & PI & \\ [0.5ex]

	(1) & (2) & (3) & (4) & (5) & (6) & (7) & (8) & (9) & (10)\\
    \hline
    F00163-1039 N & 0.17 $\times$ 0.14 & 0.15, 83 & 82 & 0.53 &\checkmark &2.1 & 64.0 & TDS  & Pa$\beta$\\
    F00163-1039 S  & 0.15 $\times$ 0.12 & 0.13, 73 & -77 & 0.53 & \checkmark& 2.1 & 38.1 & TDS & Pa$\beta$\\
    F04315-0840 & 0.22 $\times$ 0.15 &0.19, 61 & -74  & 0.43 & & 11.7 & 24.8 & MPS & Pa$\alpha$\\
    F07160-6215 & 0.24 $\times$ 0.21 &0.23, 48 & 88 & 0.51 & & 9.8 & 24.7 & MPS  & Pa$\alpha$\\
    08355-4944 & 0.13 $\times$ 0.10 &0.11, 62  & 57 & 0.48 & & 1.7 & 25.1 & TDS  & Pa$\beta$\\
    F10015-0614 & 0.26 $\times$ 0.21 & 0.24, 87 & -83 & 0.35 & & 9.6 & 24.8 & MPS  & Pa$\alpha$\\
    F10257-4339   & 0.23 $\times$ 0.21 &0.22, 48 & 63& 0.43& \checkmark & 5.8 & 47.4 &KS  & Pa$\alpha$\\
    F11506-3851 & 0.30 $\times$ 0.24 &0.27, 53 & 63 & 0.89 & & 9.1 & 24.6& LC1  & Pa$\alpha$\\
    F12596-1529 W & 0.23 $\times$ 0.17 &0.20, 74& 89 & 0.48 & \checkmark & 9.5 & 48.6 & MPS  & Pa$\alpha$\\
    F12596-1529 E & 0.23 $\times$ 0.17 &0.20, 74 & 89 & 0.48 & \checkmark & 9.5 & 48.6 & MPS  & Pa$\alpha$\\
    13120-5453  & 0.15 $\times$ 0.11 &0.13, 85  &-21 & 0.57 & &2.3 & 25.2 & TDS  & Pa$\beta$\\
    F13182+3424  & 0.26 $\times$ 0.16 &0.21, 106  & -16  & 0.43& & 3.3	& 25.0 & TDS  & Pa$\beta$\\
    F13229-2934   & 0.31 $\times$ 0.22 &0.27, 76 & 63 &  0.21 & &10.2& 24.8& LC2  & Pa$\alpha$\\
    F13373+0105 E  & 0.19 $\times$ 0.14 &0.17, 84  & -83  & 0.59 &\checkmark & 2.6	& 50.6 & TDS  & Pa$\beta$\\
    F13373+0105 W  & 0.19 $\times$ 0.14 &0.17, 84 & -83 & 0.60  & \checkmark& 2.6 & 50.6 & TDS  & Pa$\beta$\\
    F13497+0220 S  & 0.17 $\times$ 0.14 &0.16, 112  & -82  & 0.33 &\checkmark & 2.6	&  46.1 & TDS  & Pa$\beta$\\
    F13497+0220 N  & 0.17 $\times$ 0.14 &0.16, 112 & -82 & 0.38 &\checkmark & 2.6	& 38.5 & TDS  & Pa$\beta$\\
    F14544-4255 W & 0.23 $\times$ 0.20 &0.22, 76 & -86 & 0.46 & &10.7 & 24.8 & MPS  & Pa$\alpha$\\
    F14544-4255 E & 0.23 $\times$ 0.20 &0.22, 73 & -87 & 0.47 & & 10.7 & 24.8 & MPS  & Pa$\alpha$\\
    15206-6256 S & 0.18 $\times$ 0.16 &0.17, 106 &-5 & 0.37 &\checkmark & 3.0	& 38.4 & TDS  & Pa$\beta$\\
    15206-6256 N   & 0.18 $\times$ 0.16 &0.17, 106 &-5   & 0.37 & \checkmark& 3.0	& 38.4 & TDS  & Pa$\beta$\\
    F16164-0746  & 0.14 $\times$ 0.11 &0.13, 76 & -57  & 0.43 & & 2.1& 25.1 & TDS  & Pa$\beta$\\
    F17138-1017 & 0.26 $\times$ 0.22 &0.24, 87 & -62 &0.75 & & 7.8 & 24.8 & MPS  & Pa$\alpha$\\
    F18341-5732 & 0.19 $\times$ 0.16 &0.18, 62 & -60 & 0.77  & & 2.7& 24.8 & TDS  & Pa$\alpha$\\
    F21453-3511  & 0.26 $\times$ 0.22 &0.24, 87 & -62 & 0.29 & &10.5 & 24.8 & MPS  & Pa$\alpha$\\
    F22132-3705  & 0.17 $\times$ 0.16 &0.17, 42 & -61 & 0.45 & & 9.8	& 24.7 & MPS  & Pa$\alpha$\\
    F23007+0836  & 0.16 $\times$ 0.12 & 0.14, 46 & -48  & 0.29 & \checkmark & 2.9 & 38.0 & TDS  & Pa$\alpha$\\ [1ex]
       	\hline
	\end{tabular}
    
\tablefoot{Col. (1): IRAS denomination from \cite{Sanders2003}. Col. (2): major ($\theta_{maj}$) and minor ($\theta_{min}$) FWHM beam sizes. Col. (3): mean FWHM beam size ($\theta_{m}$) in arcseconds and parsecs, respectively. Col. (4): position angle (P.A.) in degrees. Col. (5): 1$\sigma$ line sensitivity per channel (width of $\sim$ 10 km/s) of the CO(2--1) observations.  Col. (6): \checkmark galaxies with mosaics. 
 Col. (7): Maximum recoverable scales.  Col. (8): Field of view. Col. (9): Principal investigator of the ALMA project: MPS: Miguel Pereira-Santaella (2017.1.00255.S),  KS: Kazimierz Sliwa (2015.1.00714.S), LC1: Luis Colina (2013.1.00271.S), LC2: Luis Colina (2013.1.00243.S) and TDS: Tanio D\'iaz-Santos (2017.1.00395.S) Col (10): H recombination line  HST images.}
	\end{threeparttable}
    \label{tab:almainfo}
\end{table*}

\subsection{HST/NICMOS and HST/WFC3 data}\label{hst:observations}
We used the continuum subtracted NIR narrow-band Pa$\alpha$ 1.87 $\mu$m and Pa$\beta$ 1.28 $\mu$m images taken with the NICMOS and WFC3 instruments, respectively, on board HST to map the distribution of recent star formation in the galaxies of the sample (see Table \ref{tab:almainfo}).

These archival HST images are drawn from two projects: Project ID: 10169 (PI: A. Alonso-Herrero) for Pa$\alpha$ HST/NICMOS images, and Project ID: 13690 (PI: T. Díaz-Santos) for Pa$\beta$ HST/WFC3 images. The galaxies from the first project were selected to have a redshift range of 0.0093 $\leq$ \textit{z} $\leq$ 0.0174, ensuring that the Pa$\alpha$ emission line lies within the HST NICMOS F190N narrowband filter \citep{AH2006}. The FoV of the images is approximately 19$\arcsec$.5$\times$19$\arcsec$.5 ($\sim$4.2--7.4 kpc). Galaxies in the Pa$\beta$ sample (second project) were selected to have redshifts in the range 0.0225 $\leq$ \textit{z} $\leq$ 0.0352, ensuring that the Pa$\beta$ emission line lies within the HST WFC3 F132N narrowband filter.
The FoV of the latter images is much larger (140$\arcsec\times$124$\arcsec$), but most of the Pa$\beta$ emission is mostly contained within 5~kpc from the galaxy nucleus.
For details on the data reduction we refer the reader to \cite{SG2022} for Pa$\alpha$ observations and \cite{Larson2020} for Pa$\beta$ observations.

To obtain the final images, we subtracted the background emission and corrected the astrometry using stars within the NICMOS FoV in the F110W ($\lambda_{eff}$ = 1.13$\mu$m) or F160W ($\lambda_{eff}$ = 1.60$\mu$m) filters and the Gaia DR3 catalogue \footnote{\href{http://www.cosmos.esa.int/web/gaia/dr3}{http://www.cosmos.esa.int/web/gaia/dr3}}. Three objects (MCG--02--33--098~E/W, and IC4518~E) do not have Gaia stars in their NICMOS image FoV. In these cases we adjusted the astrometry using the NICMOS counterparts of multiple bright structures detected in both the ALMA continuum and CO(2--1) maps. While small positional offsets between NIR and radio emission are possible, we minimised such offsets by aligning several matching features across both datasets and maximising the overlap over the entire galaxy rather than relying on a single point-like source.
 After that, the images were rotated to have the standard north-up, east-left orientation.
In order to ensure a consistent comparison between the ALMA and HST data, the Pa$\alpha$ and Pa$\beta$ images (with spatial resolutions of 25--98 pc) were convolved with a Gaussian kernel to match the resolution of the ALMA maps at $\sim$100 pc. 
Only one galaxy has HST data at resolution lower than 90 pc (NGC\,5331, where the original spatial resolution was $\sim$98 pc). In this case we convolved the data to the lower angular resolution. The bottom--right panel of Figure \ref{fig:maps} shows the final image for NGC3110
(similar figures for the rest of the sample can be found in the database on the Xtreme Scientific Project website
\footnote{\href{https://xtreme.ia.forth.gr/}{https://xtreme.ia.forth.gr/}}
\footnote{\href{http://quasar.physics.uoc.gr/xtreme-databases/C-SKDB/figures/}{http://quasar.physics.uoc.gr/xtreme-databases/C-SKDB/figures/}}).

Most of the galaxies in the Pa$\alpha$ sample are isolated galaxies and early-stage mergers; while, mid- to late- stage mergers are covered by the Pa$\beta$ sample. The combination of the Pa$\alpha$ and Pa$\beta$ samples contains 12 isolated and galaxy pairs,  17 early-stage mergers, and 17 mid- and late-stage mergers. When all the galaxies in pairs are counted, there are a total of 59 individual galaxies in the HST sample. 
From these, we have ALMA observations at $\lesssim$ 100~pc for 27 individual galaxies ($\sim$46\% of the sample), with a relatively uniform representation of the entire merger sequence: 6 isolated galaxies, 3 pairs of galaxies and 5 early- (9 individual galaxies), 4 mid- and 5 late-stage mergers.

\section{Analysis}

\subsection{Data analysis}\label{analysis}
In this section, we describe the two different methodologies used to study the star formation relations in our sample. One focuses on analysing the gas properties along unresolved lines-of-sight, or beam-sized regions, within the galaxies, while the other examines the properties of molecular clouds by identifying and selecting physical clump structures.  
We exclude from our analysis the pixels where AGNs are located by applying a mask. 

\subsubsection{Beam-sized region analysis}\label{beamanalysis} 
\begin{figure*}[ht!]

   \centering
    \includegraphics[width=.85\linewidth]{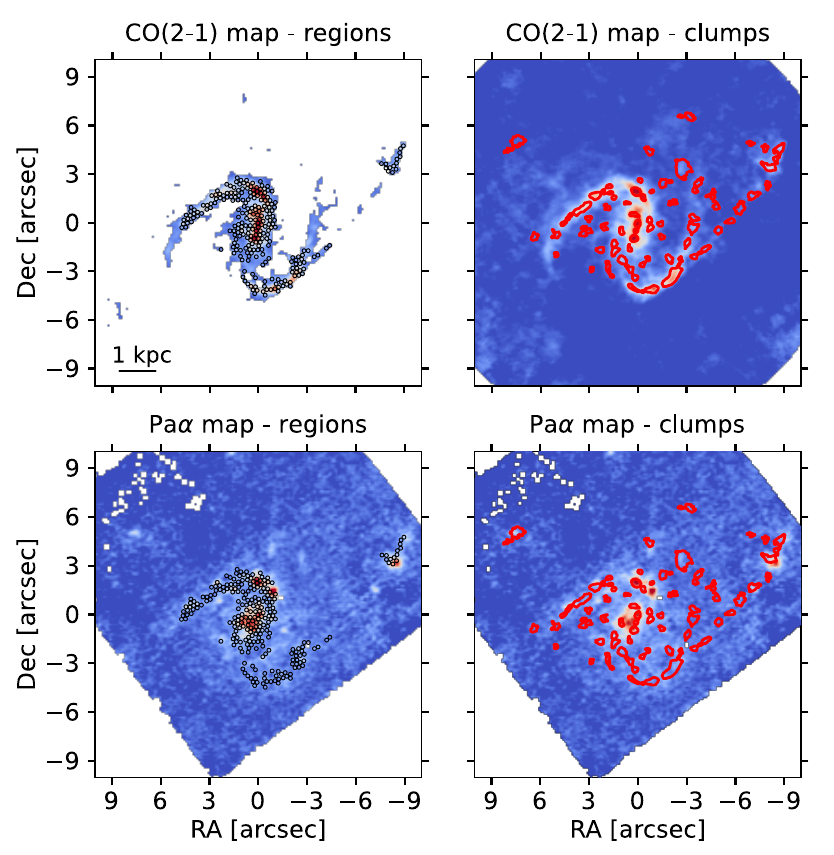}
 \caption{ALMA CO(2--1) integrated intensity (moment 0) map and HST Pa$\alpha$ image of the galaxy NGC3110. Left: Location of the beam-sized regions (in black) on the CO(2--1) (top) and Pa$\alpha$ (bottom) maps. 
 Right: Location of physical structures (in red) found using Astrodendro (clumps) on the CO(2--1) (top) map, also projected over the Pa$\alpha$ (bottom) map. The regions and clumps are both detected with ALMA and HST, considering the conditions described in Section \ref{analysis}. }
    \label{fig:methods}
\end{figure*}

To study the distribution of the cold molecular gas and the recent star formation, we define circular apertures centred on local maxima in the CO(2--1) emission maps constructed with the 3$\times$RMS$_{CO}$ procedure (see Sect. \ref{co:observations}). This method selects only the most reliable  CO emission. The apertures have a fixed diameter of $\sim$100 pc, corresponding to the spatial resolution of the data. To do
so, we first sorted the CO moment 0 pixel intensities. Then we
defined circular regions using as centre the pixels in order of
descending intensity to prevent any overlap between the regions.
With this method we obtain independent, unresolved, non-overlapping
regions centred on local emission maxima that cover all the CO
emission in each galaxy \citep[see][]{SG2022}. Once we had the regions in CO(2--1) emission maps, we used them on the Pa$\alpha$ and Pa$\beta$ maps 
to extract the associated star formation rates. We considered Pa$\alpha$ and Pa$\beta$ as detections when the line emission is above 3$\times$RMS$_{Pa}$.
The RMS$_{Pa}$ in these images corresponds to the background noise. In total, we obtain more than 4000 regions for the whole sample. The left panels of Fig. \ref{fig:methods} show an example of the location of the regions on the CO(2--1) (top panel) and Pa$\alpha$ (bottom panel) maps of the galaxy NGC\,3110 (similar figures for the rest of the sample can be found in the database on the Xtreme Scientific Project website).

\subsubsection{Dendrogram analysis}\label{sec:astrodedro}

Our data allow us to study molecular gas at scales of $\sim$100 pc, enabling the characterisation of the properties that define molecular clouds in LIRGs. The sizes of molecular clouds in galaxies are on the order of 50 to 150 pc \citep[e.g.][]{Miville2017, Oey2003, Bolatto2008}. To identify physical  clumps and measure their properties, we use the Python code \texttt{Astrodendro} \citep{Rosolowsky2008}.
This algorithm enables the study of the hierarchical structure of the CO(2--1) emission from molecular clouds. We also apply this structure to the HST recombination line images to extract the associated star formation rates. 
The dendrogram algorithms build a tree structure that is separated into three categories: leaves, branches, and trunks. The process begins with the brightest pixels in the dataset, progressively adding fainter pixels and creating small structures (referred to as leaves). If the intensities of two or more non-contiguous structures differ by more than a given threshold, they are considered separate entities. Low-density gas is represented at the bottom of the hierarchical structure in a dendrogram, the initial branch of the structure (referred to as trunk). This initial branch (trunk) connects to other branches and leaves. 
Thus, leaves are the most isolated category, with no substructure. We associate the leaves with molecular clouds/clumps in the case of the ALMA CO(2--1) maps. We applied Astrodendro to the non-masked CO(2--1) moment-0 maps, i.e., without any 3$\times$RMS$_{CO}$ clipping, to maximise the sensitivity to extended or faint emission possibly missed by a strict threshold. This choice allows us to explore diffuse gas structures that may still host active star formation.

The identification of clouds/clumps depends on three 
inputs: (i) the minimum flux value per pixel to consider in the dataset; no value lower than this will be considered in the dendrogram (\texttt{min\_value}); (ii) how significant a leaf has to be with respect to other already detected leaves or branches in order to be considered an independent structure (\texttt{min\_delta}). The significance is measured from the difference between its peak flux and the value at which it is being merged into the tree, and (iii) the minimum number of pixels (equivalent to an area) needed for a leaf to be considered an independent entity (\texttt{min\_npix}). 
We use the following conditions to identify clouds/clumps in our galaxy sample: \texttt{min\_value} = 1$\times$RMS$_{CO}$ (see below for further constraints), \texttt{min\_delta} = 1$\times$RMS$_{CO}$, \texttt{min\_npix} = 80$\%$ of a beam (PSF) area for ALMA data. Although we allow structures to be built from 1$\times$RMS$_{CO}$ emission to recover fainter regions, we impose a final constraint that all identified clumps must have an emission greater than 3$\times$RMS$_{CO}$, to avoid noise spikes. This two-step criteria ensures a conservative selection of structures while preserving sensitivity to extended or diffuse clumps.
After creating the clump catalogue based on the CO(2--1) emission maps, we measure the hydrogen recombination line emission from the Pa$\alpha$ and Pa$\beta$ maps using the same spatial coordinates and leaf structures identified in the CO(2--1) clumps. 
As for the aperture-based regions, we considered Pa$\alpha$ and Pa$\beta$ detections when the associated structures are above 3$\times\mathrm{RMS}_{Pa}$, where $\mathrm{RMS}_{Pa}$ is the background noise of each map.

Using this methodology, we identify a total of 1027 clumps in our sample of galaxies. 
The validation of this method using observations from one and two array configurations is presented in Appendix \ref{app:configurations}. We compared images obtained with a long baseline array and with a combination of long and short baseline arrays. We find consistent clump identification between both datasets.

\begin{figure}[ht!]
   \centering
    \includegraphics[width=.95\linewidth]{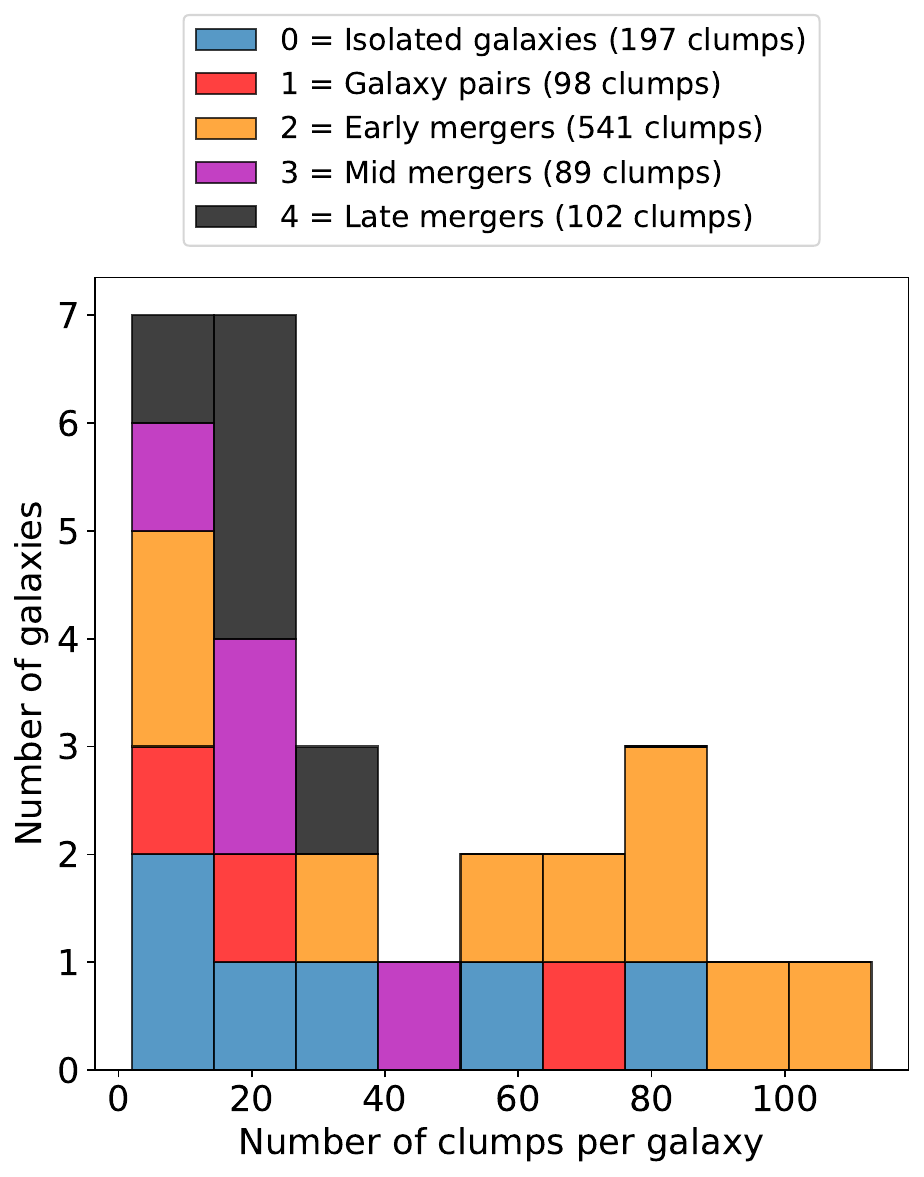}
 \caption{Distribution of the number of CO clumps per galaxy across the different merger stages in the galaxy sample. The early stages of the merger sequence cover the entire range of the stacked histogram, while galaxies in more advanced merger stages contain a lower number of clumps per galaxy.
 }
    \label{fig:histogramclumps}
\end{figure}

The right panels of Fig. \ref{fig:methods} shows an example of the location of the clumps on the CO(2--1) (top panel) and Pa$\alpha$ (bottom panel) maps of the galaxy NGC\,3110.  
Fig. \ref{fig:histogramclumps} shows the distribution of the number of gas clumps per galaxy across the different merger stages. This figure suggests that galaxies at early stages of the merger sequence (isolated galaxies, galaxy pairs and early mergers) contain from just a few dozen clumps to over 100 clumps within them, while galaxies in more advanced merger stages host comparatively lower (N<40) number of clumps per
galaxy.

\subsection{Calculating physical properties}
We measure the physical properties of each region and clump where CO and Pa$\alpha$/Pa$\beta$ are detected, following the analysis of Sect. \ref{beamanalysis} and \ref{sec:astrodedro}.
For both methods we calculate the mass, velocity dispersion of the gas, star formation rate and molecular gas surface densities, the boundedness of the gas and star formation efficiency.  
Furthermore, we estimate the effective radius
for each clump. All these parameters are detailed below. We refer to quantities computed via beam-sized region analysis as $X_{regions}$ and via dendrogram analysis as $X_{clumps}$.

To estimate the cold molecular gas mass in the beam-sized regions and clumps, we use a CO(2--1)/CO(1--0) ratio (R$_{21}$) of 0.7, obtained from single-dish CO data of the LIRG IC~4687 \citep{Albrecht2007}. This R$_{21}$ value is within the range found by \cite{Garay1993} and \cite{Arroyave2023} for infrared galaxies. We also use a Galactic CO-to-H$_{2}$ conversion factor, $\alpha^{1-0}_{CO}$ = 4.35 M$_{\odot}$pc$^{-2}$(K~km~s$^{-1}$)$^{-1}$ \citep{Bolatto2013}. 
The obtained cold molecular gas masses depend on the conversion factor ($\alpha_{CO}$) adopted. A detailed analysis about the effects of using different $\alpha_{CO}$ values on our results of this work is provided in Appendix \ref{app:conversionfactor}. 
LVG/RADEX studies of individual LIRGs often find lower, ULIRG-like $\alpha_{CO}$ values \citep[e.g.][]{Israel2009, Sliwa2013, Sliwa2014, Sliwa2017, He2020}. However, these works typically focus on central regions, interacting pairs, or kpc-scale resolution, and therefore may not be representative of the extended discs or the full diversity of merger stages in our sample. \cite{Violino2018} also report a wide range of  $\alpha_{CO}$ (0.97--7.60, median 2.29) in 11 local LIRGs observed at low resolution, further highlighting the uncertainties. We do not expect that a lower conversion factor typical of ULIRGs \citep{Papadopoulos2012} is appropriate for our targets. 
The galaxies in our sample have a mean infrared luminosity of log(L$_{IR}$/L$_{\odot}$)=11.48. 
In addition, most of the galaxies in our sample show a mean effective radius of the large scale molecular component (R$^{eff}_{CO}$) of 740 pc \citep{bellocchi2022}, while local ULIRGs show a mean value of R$^{eff}_{CO}$ = 340 pc \citep{Pereira2021}.  Therefore, it is likely that the $\alpha_{CO}$ of our sample is similar to that of normal galaxies, rather than the lower value typically observed in local ULIRGs \citep[but see][]{Saito2017, HI2019}.

The CO-to-H$_{2}$ conversion factor can also be affected by the metallicity of the galaxies, showing higher values with decreasing metallicity \citep[$\alpha_{CO}$ = 4.35~(Z/Z$_{\odot}$)$^{-1.6}$ M$_{\odot}$pc$^{-2}$(K~km~s$^{-1}$)$^{-1}$,][]{Accurso2017}. 
\cite{Rich2012} studied the metallicity in some local (U)LIRGs, showing a decrease in the abundance with increasing radius. In the case of the metallicity in local discs, \cite{Ssanchez2014} observed a similar behaviour. Based on these works, the expected variation in the conversion factor due to metallicity gradients at $r<5$~kpc is 20-30$\%$. Finally, we calculate the molecular gas mass surface density ($\Sigma_{H_{2}}$).

To determine the SFR and SFR surface density ($\Sigma_{SFR}$) of the beam-sized regions and clumps detected in our galaxies, we use the H$\alpha$ calibration \citep{Kennicutt2012}, which assumes a \cite{Kroupa2001} initial mass function, and a H$\alpha$/Pa$\alpha$ ratio of 8.6 and a H$\alpha$/Pa$\beta$ ratio of 17.6  \citep[case B at T$_{e}$ = 10.000 K and n$_{e}$ = 10$^{4}$ cm$^{-3}$,][]{Osterbrock2006}. 
To correct the Pa$\alpha$ and Pa$\beta$ emission for dust attenuation for each region and clump, we use the Br$\delta$ and Br$\gamma$ line maps observed with the SINFONI instrument on the Very Large Telescope (VLT) to derive A$_{K}$. These maps have effective FoV between 8$\arcsec\times8\arcsec$  and 12$\arcsec\times12\arcsec$ \citep{piqueras2013}. The estimated A$_{K}$ values in this work range from 0.95$\pm$0.60 to 1.98$\pm$1.29 mag. For an in-depth explanation of the extinction correction method, we refer the reader to \cite{SG2022}.

All the $\Sigma_{H_{2}}$ and $\Sigma_{SFR}$ values are corrected for the inclination of each galaxy (see Table \ref{tab:sample}, Column (7)). The inclination is estimated from Spitzer 3.6 $\mu$m band images, which have a PSF of $\sim$1.7$\arcsec$, as presented in \cite{Pereira2011}. We define the inclination ($i$) of a galaxy as
\begin{equation}
i~[^\circ] = \cos^{-1}(c_{\mathrm{sm}}/a_{\mathrm{sm}})
\end{equation}
where $c_{\text{sm}}$ and $a_{\text{sm}}$ are the lengths of the minor and major semi-axes of the object, respectively.
We use the elliptical isophote fitting from the \texttt{isophote}\footnote{\href{https://photutils.readthedocs.io/en/latest/user_guide/isophote.html}{https://photutils.readthedocs.io/en/latest/user\_guide/isophote.html}} package in Python, which provides the values of the major and minor semi-axis lengths for each fitted ellipse. Finally, the inclination value is determined as the mean inclination of the fitted ellipses that are unaffected by the PSF of the images or the structure of the galaxy. The uncertainty in the inclination is calculated as the standard deviation of the inclinations of the considered ellipses. 

We note, however, that for mid- and late-stage mergers, where galaxy morphologies are strongly disturbed and deviate from disc-like shapes, the inclination correction becomes less reliable. To mitigate this, we estimated the inclination from the most regular isophotes in the 3.6 $\mu$m images, avoiding regions strongly affected by tidal distortions or PSF effects. We also verified that our main conclusions do not change significantly when the inclination correction is omitted for these systems.

Both the SFR and the cold molecular gas surface density estimates are affected by flux calibration errors. We assume an uncertainty of about 10\% for the ALMA fluxes (see ALMA Technical Handbook \footnote{\href{http://almascience.eso.org/documents-and-tools/latest/documents-and-tools/cycle8/alma-technical-handbook}{http://almascience.eso.org/documents-and-tools/latest/documents-and-tools/cycle8/alma-technical-handbook}}), and $\sim$15--20\% for the NICMOS fluxes \citep{AH2006,Boker1999}. 

We also explore other properties of the molecular gas: the velocity dispersion ($\sigma_{v}$) obtained from the CO(2--1) moment 2, and the dynamical state of molecular gas in the beam-sized regions and clumps using the boundedness parameter 
\begin{equation}
b \,[M_\odot\,\mathrm{pc}^2\,(\mathrm{km\,s^{-1}})^{-2}] 
\equiv \Sigma_{\rm mol}/{\sigma_v^2} \propto \alpha_{\rm vir}^{-1}
\end{equation}

 where $\sigma_{v}$ is the velocity dispersion and $\alpha_{vir}$ the virial parameter. That is, low values of \textit{b} indicate that gas clumps are supported against gravity by their internal velocity dispersion, while large values of \textit{b} are representative of clumps that are unstable against gravitational collapse. 
Finally, we calculate the star formation efficiency as SFE = $\Sigma_{SFR}$/$\Sigma_{H_{2}}$ = $t_{\rm dep}^{-1}$, where $t_{\rm dep}$ is the gas depletion time.  
The effective radius of the gas clumps is calculated from the area of the clumps, as determined by the Astrodendro algorithm.

\section{Results and discussion}
In total, we define more than 4000 beam-sized regions across the whole sample (27 objects), and consider more than 1000 molecular gas clumps using Astrodendro. The sizes of these molecular gas clumps range from 89 to 694 pc, except for a clump in the galaxy IRAS15206--6256 N, which reaches 1068 pc. The median clump size in our sample is 154 pc. 
The number of clumps and regions per LIRG varies across the sample (see Figs. \ref{fig:histogramclumps} and \ref{fig:histogramregions}).

These cloud-scale observations reflect both the typical locations where stars form and 
regions with more extreme conditions than those found in star-forming galaxies.  A detailed analysis of the lifecycle of clouds will be presented in S\'anchez-Garc\'ia et al. (in prep.). In this work, we focus on  scaling relations between ISM and resolved clump properties.

\subsection{Resolved star-formation properties in LIRGs}

\subsubsection{Kennicutt-Schmidt law}
\begin{figure*}[htbp!]
   \centering   
   \includegraphics[width=14.8cm]{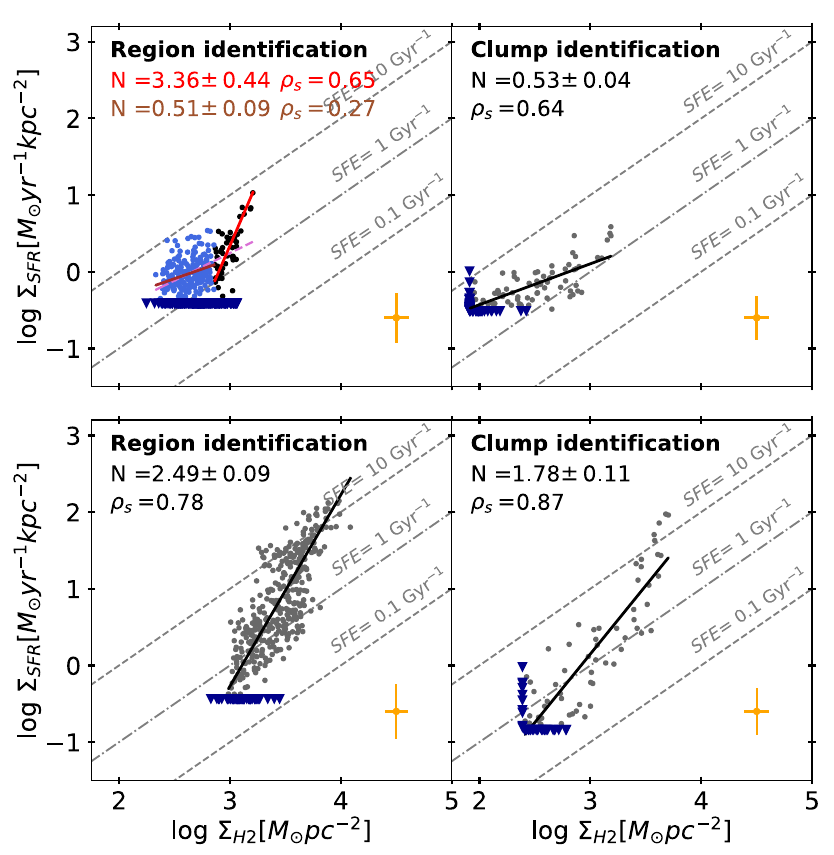}

   \caption{The SFR surface density ($\Sigma_{SFR}$) as a function of the molecular gas surface density ($\Sigma_{H2}$) 
   is shown using beam-sized regions (left) and clumps identified with Astrodendro (right) for NGC~3110 (top) and NGC~7469 (bottom). The blue and black points represent the two branches derived by applying the MARS method with breaking points in the log $\Sigma_{H2}$ for NGC3110 in the case of regions. The dark gray points correspond to the clumps method for both galaxies and the region-based method for NGC~7469. The brown and red solid lines are the best fit for the two branches, while the black line represents a one-slope fit. The Spearman’s rank correlation coefficients ($\rho_{s}$) and the power-law indices (N) of the derived best-fit KS relations in the top-left of each panel. The error bars indicate the mean systematic uncertainties in $\Sigma_{H2}$ of $\pm$ 0.11 (0.10) dex
   (horizontal) and the extinction correction in $\Sigma_{SFR}$ of $\pm$ 0.27 (0.29) dex 
   (vertical) for NGC~3110 (NGC~7469). The inverted triangles indicate upper limits. The grey dashed lines mark constant star formation efficiencies (SFE = $\Sigma_{SFR}$/$\Sigma_{H2}$).} 
              \label{fig:ksplot}
  \end{figure*}

We study the molecular KS relation for each galaxy at a resolution of $\sim$100 pc, using both beam-sized regions and clump-based measurements. 
As an example, Fig. \ref{fig:ksplot} shows the $\Sigma_{SFR}$ as a function of $\Sigma_{H2}$ for NGC~3110 and NGC~7469 (similar figures for the rest of the sample can be found in the database on the Xtreme Scientific Project website). 
The KS diagram using beam-sized regions in NGC~3110 (top left panel) suggests that these regions follow two different power laws. These two branches were identified using the Multivariate Adaptive Regression Splines \citep[MARS,][]{Friedman1991} fit, 
which gives the position of the breaking points for a linear regression with multiple slopes. The branch with higher gas and SFR densities is located in the central region of the galaxy, and shows a superlinear slope, while the other branch with lower gas and SFR densities is located in the more external disc regions, with a sublinear slope. The spatial distribution in the dual galaxy NGC~3110 is shown in Figure \ref{fig:regions_dual}, Appendix \ref{app:ksplots}. This behaviour is observed in 9 galaxies ($\sim$33\%) of our galaxy sample. These dual galaxies span the merger sequence as follows: 22\% are isolated galaxies, 12\% are galaxy pairs, and each of  early-, mid-, and late-stage mergers accounts for 22\%. Regarding their nuclear activity classification, 45\% of the dual galaxies exhibit HII region-like activity, 22\% are classified as Seyfert 2, while LINERs and composite account for 11\% each. The duality is reinforced if we consider a factor $\alpha_{CO}$ = 0.8 \citep{downes1998} typical of ULIRGs in the central regions of our sample. For the case of the clump selection (top right panel), the data points follow a single power-law. 

Similarly, the KS diagram for NGC~7469, using the beam-sized region selection (bottom left panel in Fig. \ref{fig:ksplot}), also suggests that the data points follow a single power-law. A comparable result is observed with the clump selection (bottom right panel). Thus, LIRGs exhibit two distinct behaviours when analysed using the beam-sized selection method, consistent with \cite{SG2022}, which they refer to as dual and non-dual behaviour. However, only a single behaviour is evident when the clumps are identified using Astrodendro. We fit the data points using the orthogonal distance regression \citep[ODR,][]{Boggs1987} method, which minimises the orthogonal distances from the data points to the regression line.

The absence of a dual slope (the steeper slope) in the clump-based KS relation may be related to the scarcity of clumps identified in the central kpc of galaxies that exhibit dual behaviour in the region-based KS relation. 
Such a distinction naturally appears in the region-based approach, where the central kiloparsec is treated as multiple independent apertures, but remains hidden in the clump-based method, in which the unresolved central structure contributes practically as a single clump (e.g., NGC 7130; see Fig. \ref{fig:regions_dual_2methods}). We tested several combinations of parameters in the clump-finding algorithm to evaluate whether this absence could be driven by the merging of bright structures, especially in the central regions. Even when adopting more aggressive thresholds aimed at breaking down faint structures, the central emission remained unresolved into multiple substructures. This behaviour affects only a few galaxies in our sample, and interestingly, these are precisely the systems that show a clear dual behaviour in the region-based analysis. This suggests that their central regions may host intrinsically dense and compact structures, with multiple clumps along the line of sight that cannot be separated at the current angular resolution. This scenario is consistent with the increased gas and dust compaction observed in some LIRGs by 
\cite{Tanio2017}. Higher-resolution observations are needed in order to disentangle physical gas clumps at the core of these galaxies.

As expected, these two methods show variations in both the slope and the correlation coefficient, with steeper slopes and stronger correlations observed when studying physical structures in galaxies with a non-dual behaviour. Around 70\% of the objects show a steeper slope and more than 50\% show a better correlation with the clump selection method. This result may be due to the fact that we are now studying physical structures where star formation is produced, in contrast to the region-based method, where we examine the molecular gas present in each galaxy. In the latter method, this does not imply that star formation occurs in all gas regions across the galaxy.

We do not include in our analysis upper limits. 
\cite{Pessa2021} studied the influence of non-detections in several resolved scaling relations. In principle, non-detections could artificially flatten the relations at small spatial scales, resulting in a steepening when the analysis is carried out at larger spatial scales, as pixels with signal would be averaged with the non-detection pixels at larger scales. In turn, they found that ignoring the non-detections have a small impact on the measured slope.

\cite{Zetterlund2019} studied how the properties of a set of Galactic GMCs correlate with the local SFR using Astrodendro. They found a steep slope of N=2 in the KS diagram, with a 1$\sigma$ scatter of $\sim$0.6 dex. In our sample we obtain a wide range of slopes from sub-linear (0.39$\pm$0.07) up to super-linear (3.36$\pm$0.44). 
However, \cite{Demachi2024} studied the GMCs properties using the PHANGS galaxy M74 with a resolution of 50 pc, finding a large dispersion in the relation. Their results suggested that the law breaks down at a GMC scale (<100 pc). This contrasts with the results obtained in this work, where we observe a strong correlation ($\rho_{s}$>0.7) for most of the galaxies and a significantly smaller scatter, in the range of 0.10–0.25 dex. This discrepancy could arise from the fact that in M74, the identified structures are on average smaller and may represent substructures of the GMCs, producing the large dispersion due to the different evolutionary stage of these substructures. In other words, this galaxy may not exhibit the same limitations in the identification of structures as those observed in the central regions of our sample. Another possible explanation 
for the tighter relation we observe is that, in LIRGs, most GMCs are in active star-forming stages, while in normal spirals a large fraction of GMCs remain quiescent, thereby increasing the dispersion. \cite{Kreckel2018} measured that only 22\% of the GMCs host associated SF, with 40\% of those having two or more associated HII regions in NGC~628. This could result in a wider range of dispersion on the y-axis of the KS diagram.

\subsubsection{Self gravity of the gas} \label{sect:sigma_inidividual}
\begin{figure*}[htbp!]
   \centering
 
\includegraphics[width=18.4cm]{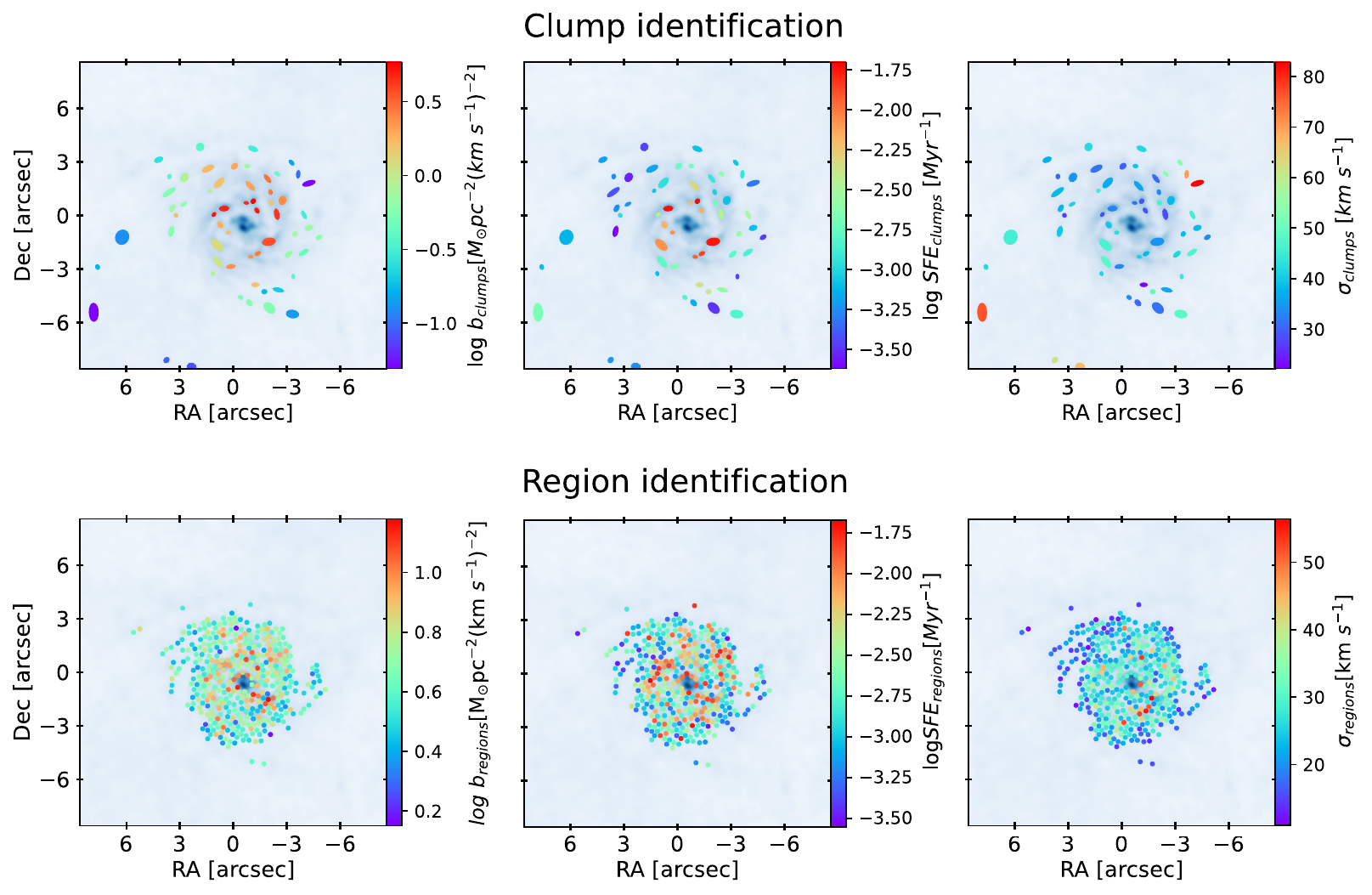}
   \caption{Maps of the boundedness parameter, $b_{clumps}$ (top left) and $b_{regions}$ (bottom left), as well as maps of the star formation efficiency, $SFE_{clumps}$ (top middle) and $SFE_{regions}$ (bottom middle), and the velocity dispersion, $\sigma_{v,clumps}$ (top right), and $\sigma_{v,regions}$ (bottom right) of the gas for the galaxy NGC~7469.
   Both methods yield comparable results, as evidenced by the spatial distribution of the regions and clumps within the galaxy.
}
     \label{fig:other_parameters_individual_galaxies}
  \end{figure*}

We explore the dynamical state of the molecular gas 
using the boundedness parameter of gas in regions ($b_{regions}$) and clumps ($b_{clumps}$). An example of the spatial distribution of the boundedness parameter, SFE ($t_{dep}^{-1}$), and velocity dispersion for one galaxy in our sample (NGC~7469) is shown in Fig. \ref{fig:other_parameters_individual_galaxies}, illustrating the type of maps used throughout the analysis. For completeness, direct scatter plots of these parameters are provided in Appendix \ref{app:parameters_plots} (Fig. \ref{fig:parameters_plots}). Equivalent figures for all galaxies are available in the Xtreme Scientific Project database. In the following, we focus on trends observed across the full sample rather than on the properties of individual systems.

The behaviour of gas boundedness varies substantially from galaxy to galaxy. In some systems, the most gravitationally bounded gas is located preferentially toward the central or dynamically disturbed regions (e.g., interacting or barred systems), while in others no clear radial pattern is present. Despite this diversity, clump-based and region-based measurements generally yield consistent results within each galaxy. When comparing the gas boundedness with the SFE on a galaxy-by-galaxy basis, we also retrieve a broad range of behaviours. Some galaxies show a positive correlation between these quantities (e.g., NGC 1614, NGC 7469), whereas others display weak or even negative correlations (e.g., IRASF 13373+0105A/B). Overall, most galaxies tend to show increasing SFE with increasing gas boundedness, although with significant scatter. This variety is consistent with the heterogeneous ISM conditions of LIRGs, which span a wide range of dynamical environments, merger stages, and gas morphologies.

Previous studies analysing the relation between gas boundedness and the depletion time ($t_{dep}$) rather than SFE; however, since both quantities are inversely related 
(SFE = $t_{dep}^{-1}$), their physical interpretation is equivalent.
\cite{Leroy2017} studied the cloud-scale ISM at a resolution of 40 pc distributed over $\sim$ 370 pc scales in the galaxy M51. They found that gas with larger 
$b$ (more bounded) exhibits shorter $t_{\rm dep}$, suggesting that increased self-gravity promotes higher star formation efficiency. 
However, \cite{Kreckel2018} did not find any correlation between $b$ and $t_{\rm dep}$ in another spiral (NGC 628) at a resolution of 50 pc over scales of 500 pc. In our work, we find different trends, indicating that, as reported in previous studies, the behaviour of self-gravity of the gas with the star formation efficiency varies from system to system. The SFE values in our sample correspond to depletion times that are 4 to 8 times shorter than those measured in the two spiral galaxies. For a lower $\alpha_{CO}$ value, we would obtain an even shorter $t_{\rm dep}$. We obtain median depletion timescales of 0.2--1.2 Gyr in our sample using a Galactic $\alpha_{CO}$ conversion factor. For the galaxy M51, depletion times vary between 1.5 and 2 Gyr, while for NGC\,628, the median depletion time ranges between 1 and 3 Gyr. This difference is consistent with what was found in previous works for starbursts \citep{daddi2010, genzel2010, Burillo2012, SG2022}.

\subsubsection{Velocity dispersion of the gas}
In this section, we explore the behaviour of the velocity dispersion of the gas in our sample. The velocity dispersion also displays considerable diversity across the sample. As illustrated in Fig. \ref{fig:other_parameters_individual_galaxies} for NGC 7469, the velocity dispersion of the gas ($\sigma_{v}$) can vary substantially within a single galaxy. Across the full sample, some systems show enhanced velocity dispersion toward their centres or in regions associated with strong dynamical perturbations, while others exhibit higher dispersion in their outer discs or no clear radial dependence. These differences likely arise from the combined influence of merger-driven turbulence, circumnuclear inflows, feedback, spiral structure, and local gravitational instabilities. These variations indicate different dynamical environments throughout the galaxy. Overall, we observe good agreement between the two methods.

The relation between velocity dispersion and SFE is likewise non-uniform. Several galaxies show positive correlations, where high-dispersion clumps tend to be more efficient at forming stars (e.g., NGC 7130, NGC 2369). Other systems, however, do not exhibit a significant trend. No universal relation is found across the sample. 
Overall, the spatial distributions and correlations involving gas boundedness, velocity dispersion, and SFE demonstrate that LIRGs encompass a wide range of ISM conditions.

\subsection{Star formation properties as a function of merger stage}
We aim to investigate the potential influence of the merger process on the star formation in galaxies. While morphological differences are well-known, this study seeks to understand whether and how the merger stage impacts star formation and the environment of molecular clouds, thereby enhancing and/or suppressing the efficiency of new star formation. 
Although both methodologies described in Section \ref{analysis} were applied across the sample, we focus here on the clump-based analysis, as it is better suited for investigating the physical properties of molecular gas structures. 
Appendix \ref{app:regions_measurements} presents the analysis based on beam-sized regions, while Appendix \ref{app:comparisons} provides a detailed comparison of the two methodologies applied across the full sample of galaxies.

\subsubsection{Kennicutt-Schimdt law across the merger sequence}

\begin{figure*}[ht]
    \centering
    \includegraphics[width=.98\linewidth]{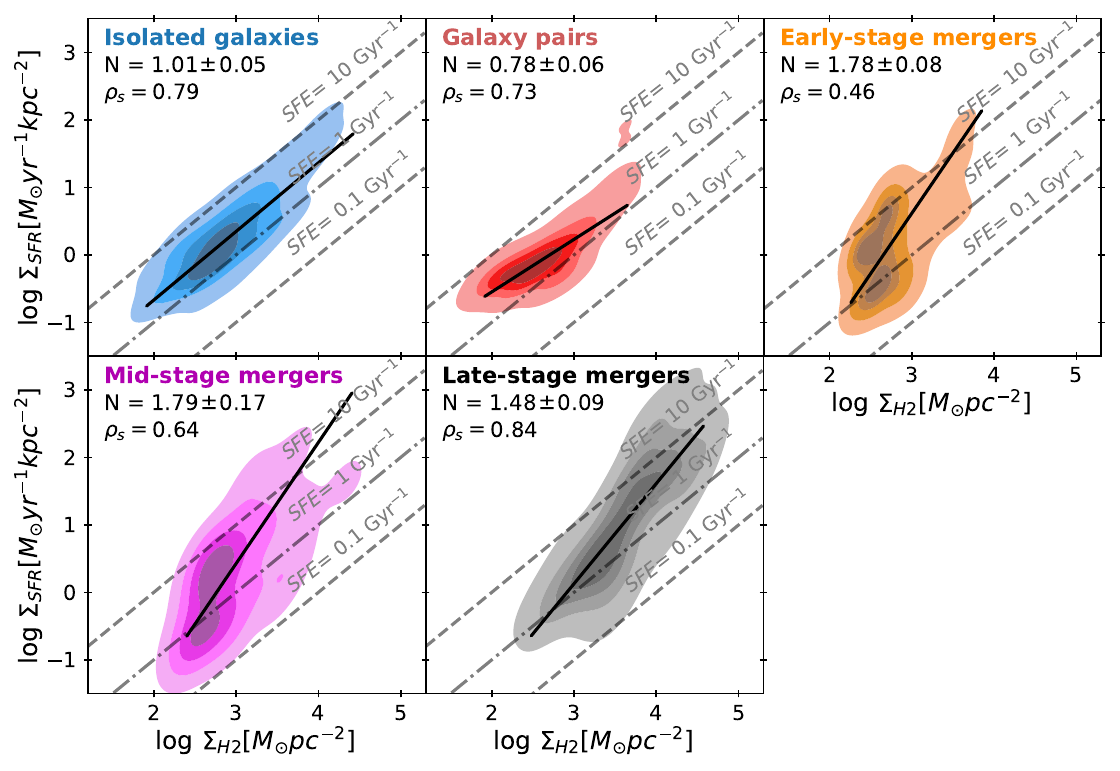}
    
    \caption{KS diagram as a function of galaxy merger stage based on clump identification. From left to right and top to bottom, the merger stages are: isolated galaxies (blue), galaxy pairs (red), early-stage mergers (orange), mid-stage mergers (magenta) and late-stage mergers (black). The black solid line represents the best fit for each dataset. The Spearman’s rank correlation coefficients ($\rho_{s}$) and the power-law indices (N) of the derived best-fit KS relations at the top-left of each panel. The grey dashed lines mark constant star formation efficiencies (SFE = $\Sigma_{SFR}$/$\Sigma_{H2}$).}
    \label{fig:ksplot_merger_sequence}
\end{figure*}

We investigate the star formation relation as a function of the merger stage of the galaxies in our sample.  
Examining the KS relation for clumps (Fig. \ref{fig:ksplot_merger_sequence}), we find that the slope of the correlation is mostly linear or sublinear in the early interacting stages of the merger sequence (isolated galaxies and galaxy pairs). However, as the merger progresses to more advanced stages, the slope becomes superlinear (steeper), implying a change in behaviour in the efficiency of converting molecular gas to stars. 
Moreover, as shown in Figure \ref{fig:ksplot_merger_sequence}, the correlations vary with the merger stage. Table \ref{tab:r2_and_meanvalues} presents the correlation coefficient, mean values, and KS relation slope for merger stages in both clumps and beam-sized regions. 

It is only when physically coherent gas structures (clumps) are identified that we observe this evolution in the KS slope, highlighting the importance of focusing on clumps to study star formation efficiency. For completeness, we also tested the KS relation using beam-sized regions; these results are presented in Appendix \ref{app:regions_measurements}.
In contrast, the uniform slope observed using the beam-sized regions method likely reflects the influence of the larger gas disc across the full extent of the galaxy, as these unresolved lines-of-sight include emission regions across the entire gas structure of the galaxies that are much more diffuse than those probed by the physical clumps. 
These results are consistent with high-resolution hydrodynamic simulations that investigate how galaxy mergers affect the structure of the ISM and the properties of GMCs and young massive clusters \citep{li2022}, as well as simulations that study the properties of young star clusters formed under different ISM conditions within a galaxy \citep{fire2022}.

\begin{figure}[ht]
   \centering
   \includegraphics[width=9cm]{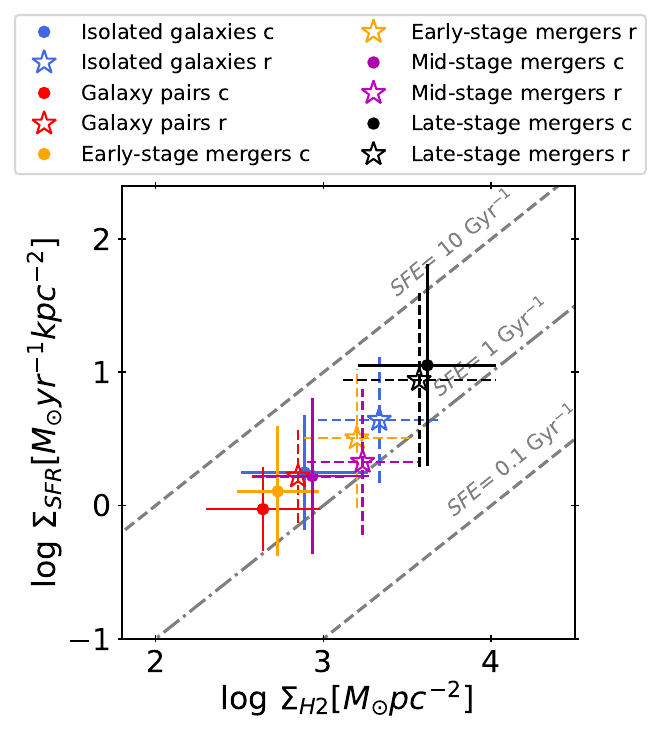}
     
    \caption{
    The mean values of the KS diagram for each merger stage using both methodologies: open stars for beam-sized regions method and dots for the clumps method. The error bars indicate the mean absolute deviation.  
    }
    \label{fig:ksplot_2methods}
    \end{figure}

The mean values of log$_{10}\Sigma_{H2}$ and log$_{10}\Sigma_{SFR}$ for each merger stage, calculated using the two different methods described in this work, show differences (see Fig. \ref{fig:ksplot_2methods}).
 For isolated galaxies, galaxy pairs, and early-stage mergers, the mean values obtained through the region selection method are slightly higher than those derived from the clump-based method (see Table \ref{tab:r2_and_meanvalues}). 
For late-stage mergers, however, both methods (clump and region selection) yield similar results. 
When focusing on the clump-based method, we find a gap of more than 0.5 dex between late-stage mergers and earlier stages.
These results indicate clear variations across the merger sequence, with higher values of $\Sigma_{SFR}$ and $\Sigma_{H2}$ in late-stage mergers compared to earlier interaction stages. 
Such differences likely reflect the progression of the merger process itself, where in isolated galaxies and early interactions the gas is more widely distributed, leading to lower surface densities of gas and star formation. As the merger advances, however, the gas in the central kiloparsecs of the merging galaxy becomes more concentrated, along with an increase in the SFR.

\begin{table*}[ht]
    \centering
    \small
        \caption{Statistical parameters of the KS diagram across the merger sequence for clumps and beam-sized regions}
        \begin{tabular}{lc cccc c cccc}
        \hline
        \hline
        Merger   & \multicolumn{4}{c}{Clumps} & & \multicolumn{4}{c}{Beam-sized regions}\\ 
        \cline{2-5} \cline{7-10}
        
         stages & $\rho_{s}$  & log$_{10}\dfrac{\Sigma_{H2}}{M_{\odot}pc^{-2}}$ & log$_{10}\dfrac{\Sigma_{SFR}}{M_{\odot}yr^{-1}kpc^{-2}}$ & Slope &  &$\rho_{s}$ & log$_{10}\dfrac{\Sigma_{H2}}{M_{\odot}pc^{-2}}$ & log$_{10}\dfrac{\Sigma_{SFR}}{M_{\odot}yr^{-1}kpc^{-2}}$ & Slope\\ 
        [0.5ex] 
        \hline
        Isolated & 0.79  & 2.89 $\pm$ 0.38& 0.25 $\pm$ 0.43 & 1.01 $\pm$ 0.05 & & 0.70  & 3.33 $\pm$ 0.37& 0.64 $\pm$ 0.47  & 0.92 $\pm$ 0.03\\
        Pairs & 0.73 & 2.64 $\pm$ 0.34 & -0.03 $\pm$ 0.32& 0.78 $\pm$ 0.06 & & 0.73 &2.85 $\pm$ 0.27& 0.22 $\pm$ 0.35& 1.05 $\pm$ 0.04\\
        Early-stage & 0.46 & 2.73 $\pm$ 0.24 &0.11 $\pm$ 0.49 & 1.78 $\pm$ 0.08 & & 0.64 & 3.20 $\pm$ 0.31& 0.51 $\pm$ 0.52& 1.09 $\pm$ 0.03\\ 
        Mid-stage & 0.64  & 2.93 $\pm$ 0.34 &0.22 $\pm$ 0.59 & 1.79 $\pm$ 0.17 & & 0.69 &3.23 $\pm$ 0.33& 0.33 $\pm$ 0.55 & 1.08 $\pm$  0.04\\
        Late-stage & 0.84  & 3.62 $\pm$ 0.41 & 1.05 $\pm$ 0.76 & 1.48 $\pm$  0.09 & & 0.85  & 3.57 $\pm$ 0.45 & 0.94 $\pm$ 0.65 &  1.05 $\pm$  0.02\\

        \hline
        \end{tabular}
        
        \tablefoot{For each merger stage (isolated galaxies, galaxy pairs, early-stage mergers, mid-stage mergers, and late-stage mergers), the table lists the Spearman’s rank correlation coefficient $\rho_{sp}$ (two-sided p-values), the mean and error bars (from the mean absolute deviation) of log$_{10}\Sigma_{H2}$ [$M_{\odot}pc^{-2}$] and log$_{10}\Sigma_{SFR}$ [$M_{\odot}yr^{-1}kpc^{-2}$], and the best-fit power-law indices. The slope of the KS law, when using clump identification, becomes super-linear in early and more advanced stage mergers, indicating larger star formation efficiencies at higher gas surface densities. 
        }
        \label{tab:r2_and_meanvalues}
\end{table*}

\subsubsection{Self gravity across the merger sequence}
\begin{figure*}[ht]
    \centering
\includegraphics[width=.98\linewidth]{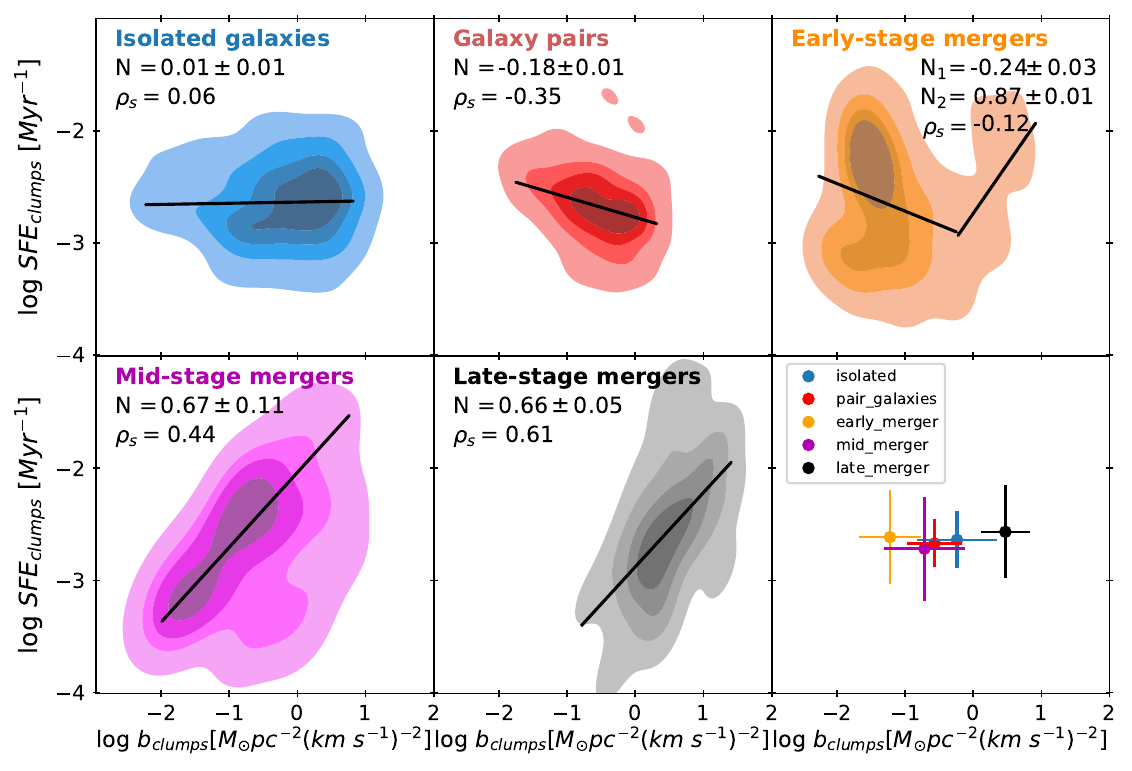}

    \caption{Star formation efficiency, $SFE_{clumps}$, as a function of the self-gravity of the clumps (parameter $b_{clumps}$) across the merger sequence. From left to right and top to bottom, the merger stages are: isolated galaxies (blue), galaxy pairs (red), early-stage mergers (orange), mid-stage mergers (magenta) and late-stage mergers (black). The Spearman’s rank correlation coefficients ($\rho_{s}$) and the power-law indices (N) of the best-fit relations are indicated.  The bottom right panel shows the mean values of $log_{10}SFE_{clumps}$ and $log_{10}b_{clumps}$ for each merger stage. The error bars indicate the mean absolute deviation.  The slope of the correlation between SFE$_{ clumps}$ and $b_{clumps}$ transitions from relatively flat, to a broken power law at early merger stages, to a positive slope in mid- and late-stage mergers, 
    showing that, at later interaction stages, the most bounded gas clumps convert their gas into stars more efficiently.
    }
    \label{fig:boundedness_means}
\end{figure*}

Figure \ref{fig:boundedness_means} shows the star formation efficiency, SFE, as a function of the self-gravity of the clumps given by the $b$ parameter. We observe a shift in the trend from negative to positive correlations as the merger progresses, with the strength of the correlation increasing at the later stages of interaction. This indicates that, in later merger stages, the self-gravity of the clumps generally have high values (most show $b_{clumps}$ $\gtrsim$ 1), and those that are more bound are also more efficient at forming stars. 
The study of the self-gravity of the gas in individual galaxies 
reveals a variety of results (see Sect. \ref{sect:sigma_inidividual}).
However, the picture changes when considering the merger stage of galaxies, and some trends emerge. 

The average SFE across the merger stages does not change significantly, but the boundedness and its relation with SFE evolve strongly (see Table \ref{tab:r2_and_meanvaluesb} and the bottom-right panel of Figure \ref{fig:boundedness_means}).
For comparison, the results obtained with the beam-sized region method show generally more uniform trends (see Appendix \ref{app:regions_measurements}, Figure \ref{fig:b_merger_sequence_regions}), reflecting the averaging over the entire galaxy rather than focusing on physical structures. This brief mention highlights the complementarity of the two methods, while the detailed comparison is provided in the appendix.

\begin{table*}[ht]
    \centering

        \caption{Statistical parameters of the SFE vs. $b$ relation across the merger sequence for clumps and beam-sized regions.}
        \resizebox{\textwidth}{!}{
        \begin{tabular}{lc cccc c cccc}
        \hline
        \hline
        Merger  & \multicolumn{4}{c}{Clumps} & &  \multicolumn{4}{c}{Beam-sized regions}\\ 
        \cline{2-5} \cline{7-10}
        
          stages &  $\rho_{s}$ & $\log_{10}\frac{b}{M_\odot\,\mathrm{pc}^{-2}\,(\mathrm{km\,s^{-1}})^{-2}}$
  & $\log_{10}\frac{\mathrm{SFE}}{\mathrm{Myr}^{-1}}$
 & slope & & $\rho_{s}$ & $\log_{10}\frac{b}{M_\odot\,\mathrm{pc}^{-2}\,(\mathrm{km\,s^{-1}})^{-2}}$  & $\log_{10}\frac{\mathrm{SFE}}{\mathrm{Myr}^{-1}}$
 & slope\\ 
        [0.5ex] 
        \hline
        Isolated galaxies & 0.06  & -0.24 $\pm$ 0.58 & -2.64 $\pm$ 0.25 & 0.01 $\pm$ 0.01 & & 0.20  & 0.46 $\pm$ 0.32& -2.93 $\pm$ 0.32 &  0.21 $\pm$ 0.01\\
        Galaxy pairs & -0.35 & -0.57 $\pm$ 0.40 & -2.67 $\pm$ 0.21&  -0.18 $\pm$ 0.01 & & -0.02 &-0.07 $\pm$ 0.20& -2.81 $\pm$ 0.22 & -0.02 $\pm$ 0.01\\
        Early-stage & -0.12 & -1.22 $\pm$ 0.46 & -2.61 $\pm$ 0.41 &  -0.24 $\pm$ 0.03 / 0.87 $\pm$ 0.01  & & 0.20 &0.33 $\pm$ 0.30 & -2.97 $\pm$ 0.41 & 0.33 $\pm$ 0.05\\ 
        
        Mid-stage & 0.44  & -0.72 $\pm$ 0.61 &-2.72 $\pm$ 0.46 & 0.67 $\pm$ 0.11& & 0.27 &0.05 $\pm$ 0.26 & -3.21 $\pm$ 0.35 & 0.41 $\pm$ 0.03\\
        Late-stage & 0.61  & 0.47 $\pm$ 0.36 & -2.57 $\pm$ 0.41 & 0.66 $\pm$ 0.05 & & -0.10  & 0.59 $\pm$ 0.44 & -3.09 $\pm$ 0.30 & -0.09 $\pm$ 0.01\\

        \hline
        \end{tabular}}
    
        \tablefoot{For each merger stage (isolated galaxies, galaxy pairs, early-stage mergers, mid-stage mergers, and late-stage mergers), the table lists the Spearman’s rank correlation coefficient $\rho_{sp}$ (two-sided p-values), the mean and error bars (from the mean absolute deviation) of log$_{10}b$ [M$_{\odot}$pc$^{-2}$(km s$^{-1}$)$^{-2}$] and log$_{10}SFE$ [Myr$^{-1}$], and the best-fit power-law indices (slopes). For early-stage mergers we use MARS fit.}
        \label{tab:r2_and_meanvaluesb}
\end{table*}

\subsubsection{Velocity dispersion of the gas across the merger sequence} \label{velocityd}

Several SF models suggest that the dynamical state of the cloud, and not only its density, affects its ability to collapse and form stars \citep[e.g.][]{Krumholz2005, Hennebelle2011, Federrath2013}. In these turbulence-regulated models, supersonic and compressive turbulence promotes the formation of dense structures where stars can form, whereas solenoidal turbulence or excessive kinetic support can reduce the efficiency. In this framework, one may expect the SFE to correlate with the level and nature of the gas velocity dispersion 
\citep{Orkisz2017}. 

\begin{figure*}[ht]
    \centering
\includegraphics[width=.98\linewidth]{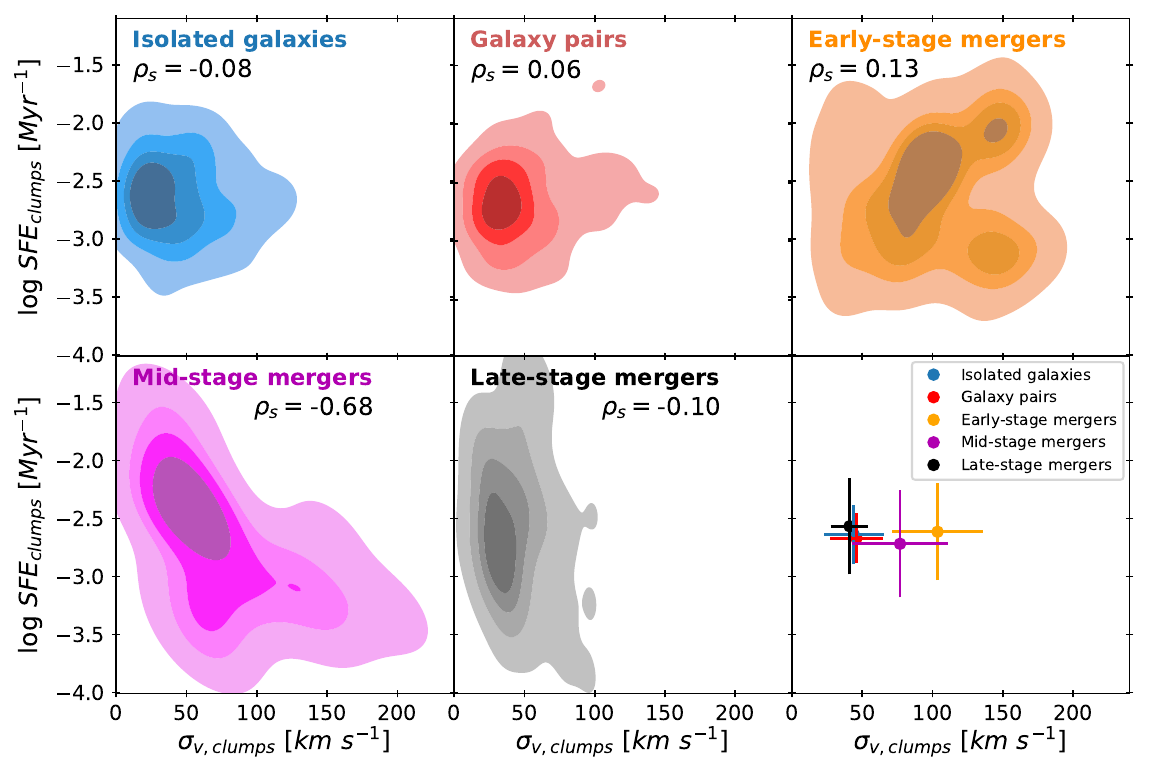}
    \caption{SF efficiency as a function of the velocity dispersion of the clumps ($\sigma_{v,clumps}$) across the merger sequence. From left to right and top to bottom, the merger stages are: isolated galaxies (blue), galaxy pairs (red), early-stage mergers (orange), mid-stage mergers (magenta) and late-stage mergers (black). The Spearman’s rank correlation coefficients ($\rho_{s}$) are indicated. The bottom right panel shows the mean values of log$_{10}SFE_{clumps}$ and $\sigma_{v,clumps}$ for each merger stage, with error bars representing the mean absolute deviation. 
    }
    \label{fig:turbulence_merger_sequence1}
\end{figure*}

Figure \ref{fig:turbulence_merger_sequence1} shows the SFE as a function of the velocity dispersion of the clumps ($\sigma_{v,clumps}$) across the merger sequence. We observe that clumps in galaxy pairs, early-stage mergers, and specially mid-stage mergers exhibit higher velocity dispersion than those in isolated galaxies and late-stage mergers.  
However, early-stage mergers exhibit significant scatter along both axes of the diagram. Additionally, we observe that clumps with high velocity dispersion are not necessarily more efficient at forming stars. In mid-stage mergers, the SFE decreases with increasing $\sigma_{v, clumps}$, 
contrary to theoretical expectations for compressive turbulence \citep[e.g.][]{Orkisz2017}.
In contrast, this anticorrelation is not observed in other merger stages (see Figure \ref{fig:turbulence_merger_sequence1}). This could be attributed to the presence of shocks resulting from the active merger, where elevated turbulence, likely driven by galaxy-wide shocks, reduces the efficiency of star formation during this chaotic stage \citep[see][]{Saito2017}, as the gas has not yet settled down. In late-stage mergers, clumps exhibit the lowest velocity dispersion among all merger stages, with a very narrow range of values. Still, the range of SFE is similar to that of mid-stage mergers. For completeness, we also examined the velocity dispersion of beam-sized regions across the merger sequence (Appendix \ref{app:regions_measurements}, Figure \ref{fig:turbulence_merger_sequence}). The beam-sized regions show more uniform trends, reflecting the contribution of the overall disc dynamics rather than the localised turbulence in clumps.

\begin{table*}[ht!]
    \centering
        \caption{Statistical parameters of the SFE vs. $\sigma_{v}$ relation across the merger sequence for clumps and beam-sized regions.}
        \begin{tabular}{lc ccc c ccc}
        \hline
        \hline
        Merger   & \multicolumn{3}{c}{Clumps} & &  \multicolumn{3}{c}{Beam-sized regions}\\ 
        \cline{2-4} \cline{6-8}
        
         stages &  $\rho_{s}$ & $\frac{\sigma_{v}}{km s^{-1}}$ & log$_{10}\frac{SFE}{Myr^{-1}}$ &   & $\rho_{s}$  & $\frac{\sigma_{v}}{km s^{-1}}$ & log$_{10}\frac{SFE}{Myr^{-1}}$ \\ 
        [0.5ex] 
        \hline
        Isolated galaxies & -0.08  & 44.00 $\pm$ 21.11& -2.64 $\pm$ 0.25 & & -0.43  & 27.72 $\pm$ 13.57& -2.93 $\pm$ 0.32 \\
        Galaxy pairs & 0.06 & 45.84 $\pm$ 18.71 & -2.67 $\pm$ 0.21&  & -0.38 &28.78 $\pm$ 10.12& -2.81 $\pm$ 0.22\\
        Early-stage & 0.13 & 103.50 $\pm$ 32.37 &-2.61 $\pm$ 0.41 &  & -0.29 & 31.13 $\pm$ 13.27& -2.97 $\pm$ 0.41\\ 
        Mid-stage & -0.68  & 76.83 $\pm$ 34.43 &-2.72 $\pm$ 0.46 & & -0.41 &43.21 $\pm$ 14.40& -3.21 $\pm$ 0.35 \\
        Late-stage & -0.10  & 40.93 $\pm$ 13.42 & 2.57 $\pm$ 0.41 & & -0.17  & 34.58 $\pm$ 12.43 & -3.09 $\pm$ 0.30 \\

        \hline
        \end{tabular}
        
        \tablefoot{
        For each merger stage (isolated galaxies, galaxy pairs, early-stage mergers, mid-stage mergers, and late-stage mergers), the table lists the Spearman’s rank correlation coefficient $\rho_{sp}$ (two-sided p-values), the mean and error bars (from the mean absolute deviation) of $\sigma_{v}$ [km s$^{-1}$] and log$_{10}$ SFE [Myr$^{-1}$].
        }
        \label{tab:r2_and_meanvaluesdv}

\end{table*}
 
\subsubsection{Interpretation}

A possible explanation for the observed trends across the merger sequence in SFE ($t_{dep}^{-1}$), $b$ parameter, and velocity dispersion is that in isolated galaxies or galaxy pairs, the influence of the interaction is still minimal or negligible. In these systems, clumps exhibit high boundedness of the gas and/or elevated velocity dispersion, but no clear correlation is observed between $t_{dep}$ and self-gravity of the gas (Fig. \ref{fig:boundedness_means}) or between SFE and velocity dispersion (Fig. \ref{fig:turbulence_merger_sequence1}). 
However, as the interaction between galaxies becomes more pronounced, tidal forces and merger-driven dynamics strongly perturb the gas, increasing turbulence and velocity dispersion within clumps. This enhances the pressure and promotes the formation of more bound GMCs, particularly in the central regions of the merging galaxies.  
Late-stage mergers, while showing moderate velocity dispersions (Fig. \ref{fig:turbulence_merger_sequence1}), still host large reservoirs of molecular gas (Fig. \ref{fig:ksplot_2methods}), allowing clumps to maintain high star formation efficiencies. 

This scenario is supported by SMUGGLE simulations \citep{li2022}, which show that stronger tidal fields increase cluster mass and mass-weighted gas pressure. This leads to a higher fraction of bound GMCs and, consequently, elevated SFE. 
Consistently, in our observations, clumps with high boundedness, high molecular gas content, and elevated pressure are preferentially located in the central regions of the mergers, coinciding with the steepest KS slopes and highest SFE. In contrast, the outskirts of these systems display lower boundedness and SFE, explaining the spatial variation in the star formation efficiency across the galaxies. 

\subsection{Radial distribution of the physical properties of clumps as a function of the merger stage}

\begin{figure*}[ht]

 \includegraphics[width=0.94\textwidth]{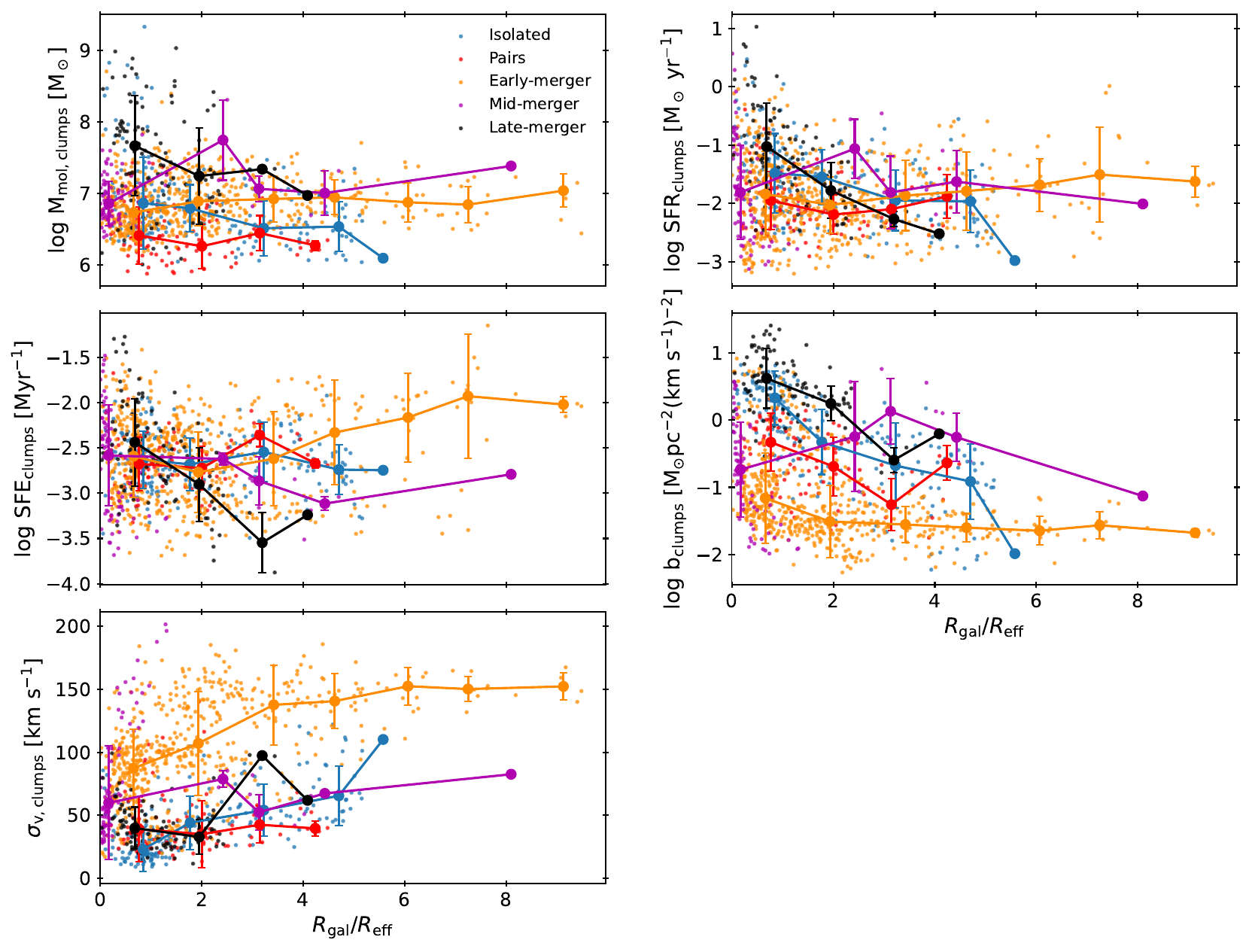}   
 
    \caption{Molecular gas mass (top left), SFR (top right), SFE (middle left), boundedness of the gas (middle right), velocity dispersion (bottom left) of individual clumps as a function of galactocentric radius normalised by the CO(2--1) effective radius, R$_{gal}$/R$_{eff}$, across the sample. The data points show the median values of M$_{mol}$, SFR, SFE, $b$ and $\sigma_{v}$ in bins of normalised galactocentric distance for each merger stage. The error bars indicate the mean absolute deviation of the points in the bins. 
    }
    \label{fig:mix_distance}
\end{figure*}

\begin{figure*}[ht!]
    \includegraphics[width=0.98\linewidth]{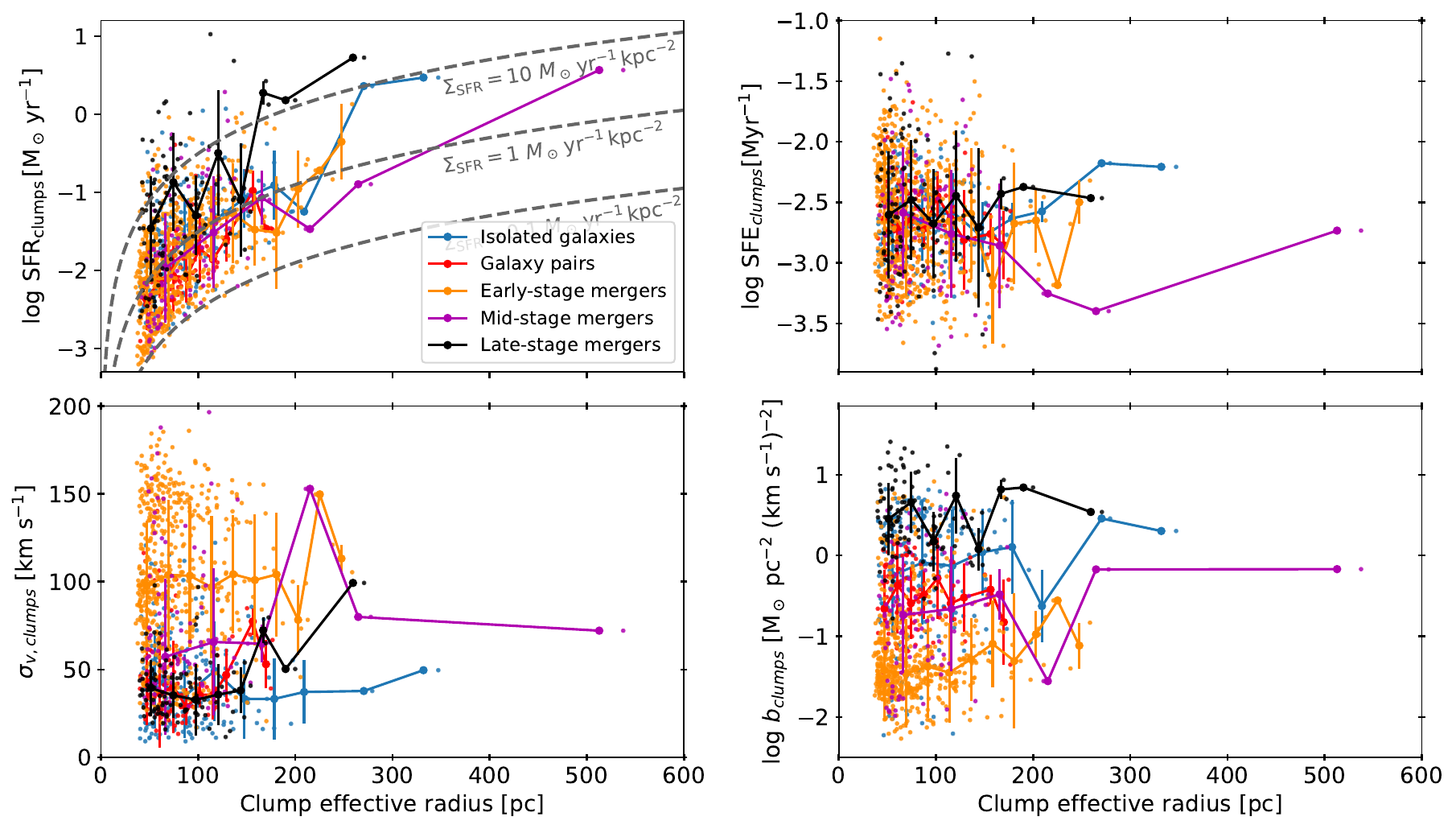}

    \caption{SFR (top left), SFE (top right), velocity dispersion (bottom left) and self-gravity (bottom right) of the clumps as function of clump sizes. The circles show the median SFR, SFE, $\sigma_{v}$ and $b$ in bins of clump size for each merger stage. The error bars indicate the mean absolute deviation of the points in the bins. In the top-left panel, the grey dashed lines indicate constant SFR surface densities.
    }

    \label{fig:mix_clump}
\end{figure*}

In this section, we analyse the variation of the gas clump properties as a function of the galactocentric distance normalised by the galactic effective radius, and also as a function of the effective radius of the clumps. To do so, we only use the results obtained through the clump method, focusing on individual clump SFRs, molecular gas masses, SFEs, velocity dispersions and their boundedness (see Figures \ref{fig:mix_distance} and \ref{fig:mix_clump}).

When examining the results as a function of merger stage in Figure \ref{fig:mix_distance}, we find that the SFR of individual clumps decreases with increasing normalised galactocentric radius, particularly for isolated galaxies and late-stage mergers, while the gas mass of the clumps remains roughly constant. 
The boundedness parameter decreases with radius across the different merger stages, except in early-stage mergers, where it remains approximately constant and at lower values compared to the others. 
The velocity dispersion of the clumps tends to increase with the normalised galactocentric radius. 

When we study the relation between clump size and various parameters (see Figure \ref{fig:mix_clump}), we find that the SFR increases with the size of the clumps, while the SFE and the self-gravity of the clumps remain approximately constant. 
The gas velocity dispersion increases with clump size in galaxy pairs, mid-stage, and late-stage mergers, while the other merger stages show no significant trend.

\cite{Larson2020} analysed star-forming regions in local (U)LIRGs from the GOALS sample, and found a clear correlation between the SFR and size of the star-forming regions, similar to that observed in high-redshift galaxies (see their Fig. 9). In our analysis, we find a qualitatively similar increase in SFR with clump size when considering molecular gas clumps and associating the SFR with the corresponding star-forming regions. However, the SFE in our sample remains approximately constant with size, indicating that the rise in SFR toward larger clumps is primarily driven by their higher gas content rather than an intrinsic increase in efficiency. Additionally, the distribution of SFR and clump size varies between merger stages, suggesting that dynamical evolution influences how gas is assembled into larger star-forming structures.

Observations of molecular clouds in our Galaxy and a number of nearby galaxies have identified various empirical trends manifesting such cloud–environment correlations. Within a galaxy, molecular clouds located closer to the galaxy centre appear denser, more massive, and more turbulent 
(e.g., \citealt{Oka2001}; \citealt{Colombo2014}; \citealt{Freeman2017}; \citealt{Hirota2018}; \citealt{Brunetti2021}). An exception to this trend is observed in the early-stage LIRG NGC 5258 (Arp 240), where the brightest clumps are entirely extranuclear \citep{saravia2025}. In our sample, we find more massive clumps and with higher SFR in late-stage mergers. 

\section{Summary and Conclusions}
We have presented a spatially resolved study of the star formation relations in a sample of 27 local LIRGs, on scales of $\sim$100 pc, spanning the entire merger sequence, from isolated to late stage mergers. We combined HST Pa$\alpha$ and Pa$\beta$ emission with ALMA CO(2--1) data to determine the SFR and molecular gas content in our galaxy sample. We have also investigated the potential influence of the merger process on the star formation in our sample, measuring the star formation efficiency, as well as the self-gravity and velocity dispersion of the gas. 
We used two different methods: 1) sampling the emission of the galaxies with beam-sized regions, and 2) identifying physical structures (clumps) using the identification algorithm Astrodendro.

The main results of this paper are summarised as follows:

\begin{itemize}
    \item We derived spatially resolved KS relations for each LIRG of the sample, using beam-sized regions and clumps. When using beam-size regions, we identified two different behaviours in the KS plot: 67\% of galaxies follow a single trend, while the remaining ones display two branches, suggesting the existence of a duality in this relation. However, when analysing physical clumps, the duality disappears, and only one single trend is observed.
    
    \item The two methods reveal variations in both the slope and the correlation coefficient. These results provide two different perspectives for studying the star formation relation: one focusing the statistics of gas and SF properties at the smallest scales, and the other targeting physically coherent gas clumps and structures.

    \item  We studied galaxies according to their merger stage. We find the slope of the KS relation becomes steeper as the merger progresses, reflecting changes in the SF efficiency of molecular gas clumps.  In late--stage mergers, we observe higher values of $\Sigma_{SFR}$ and $\Sigma_{H2}$ compared to the other stages.  However, when analysing the KS relation using beam-sized regions, the slopes remain close to one, displaying more homogeneous behaviour across the merger sequence.

    \item In isolated galaxies and up to early stage mergers, the star formation efficiency of the clumps, SFE$_{clumps}$, does not depend on their self-gravity, $b_{clumps}$. However, in later merger stages, clumps with higher boundedness become more efficient at forming stars, exhibiting higher star formation efficiencies. 
    
    \item In early- and mid-stage mergers, the clumps exhibit higher velocity dispersion compared to the other stages of the merger sequence. However, clumps in early-stage mergers exhibit higher velocity dispersion but they are not very efficient. In late-stage mergers, the velocity dispersion of the gas shows a smaller range of values.
    
\item The SFR of individual clumps decreases with increasing galactocentric distance when normalised by the effective radius, particularly in isolated galaxies and late-stage mergers, while the molecular gas mass remains roughly constant with radius. The velocity dispersion of the clumps tends to increase with the normalised galactocentric distance. The self-gravity of the gas decreases with radius across the different merger stages, except in early-stage mergers, where it remains approximately constant and lower than in the other stages. 

\item We also find that the SFR increases with the size of the clumps, although the strength of this correlation varies with the merger stage, being more pronounced in mid- and late-stage mergers. The SFE remains approximately constant with size, while the gas velocity dispersion increases with clump size in galaxy pairs, mid-stage, and late-stage mergers.

\end{itemize}

\section*{Data availability}

The catalogue of molecular gas clumps identified in this work,  including their main physical properties, is provided as Table~6. Table~6 is only available in electronic form at the CDS via anonymous ftp to cdsarc.u-strasbg.fr (130.79.128.5) or via \url{http://cdsweb.u-strasbg.fr/cgi-bin/qcat?J/A+A/}.

\begin{acknowledgements}
 We thank the anonymous referee for comments and suggestions that helped improve this manuscript.     
 MSG acknowledges that this research project  was supported by the Hellenic Foundation for Research and Innovation (HFRI) under the "2nd Call for HFRI Research Projects to support Faculty Members \& Researchers" (Project Number: 03382). MSG also acknowledges support from the National Radio Astronomy Observatory Visitor Program.  
 MSG and YS acknowledge support from the Joint ALMA Observatory Visitor Program.  
 MPS acknowledges support under grants RYC2021-033094-I, CNS2023-145506 and PID2023-146667NB-I00 funded by MCIN/AEI/10.13039/501100011033 and the European Union NextGenerationEU/PRTR. CR acknowledges support from SNSF Consolidator grant F01$-$13252, Fondecyt Regular grant 1230345, ANID BASAL project FB210003 and the China-Chile joint research fund.
 This research made use of Astrodendro, a Python package for computing dendrograms of astronomical data (\href{http://www.dendrograms.org/}{http://www.dendrograms.org/}).
 This paper makes use of the following ALMA data: ADS/JAO.ALMA\#2013.1.00243.S, ADS/JAO.ALMA\#2013.1.00271.S, ADS/JAO.ALMA\#2015.1.00714.S, ADS/JAO.ALMA\#2017.1.00255.S, ADS/JAO.ALMA\#2017.1.00395.S. ALMA is a partnership of ESO (representing its member states), NSF (USA) and NINS (Japan), together with NRC (Canada), NSTC and ASIAA (Taiwan), and KASI (Republic of Korea), in cooperation with the Republic of Chile. The Joint ALMA Observatory is operated by ESO, AUI/NRAO and NAOJ.

\end{acknowledgements}

\bibliographystyle{aa}
\bibliography{biblio2}

\begin{appendix} 
\section{Comparison between one and two configurations in ALMA data}\label{app:configurations}

From our sample of 23 LIRG
systems, 43\% (10 systems) have observations using a single 12~m configuration. The remaining 57\% (13 systems) were also studied in \cite{SG2022}, where observations were conducted with two 12~m configurations (one more extended and one more compact). The use of only one extended configuration filters out spatial scales larger than those sampled by the shortest baselines. Additionally, it is well-known that flux is lost even before reaching the maximum recoverable scale (MRS).

We are interested in determining the properties of molecular gas at scales of $\sim$ 100 pc ($\sim$0.2'') in local LIRGs. 
This scale is smaller than the MRS, so the missing flux due to the absence of short spacing is expected to be minimal. However, we expect to lose flux at scales larger than $\sim$2.5" ($\sim$ 1.2 kpc).

\begin{figure*}[t]
   \centering
   \includegraphics[width=18.2cm]{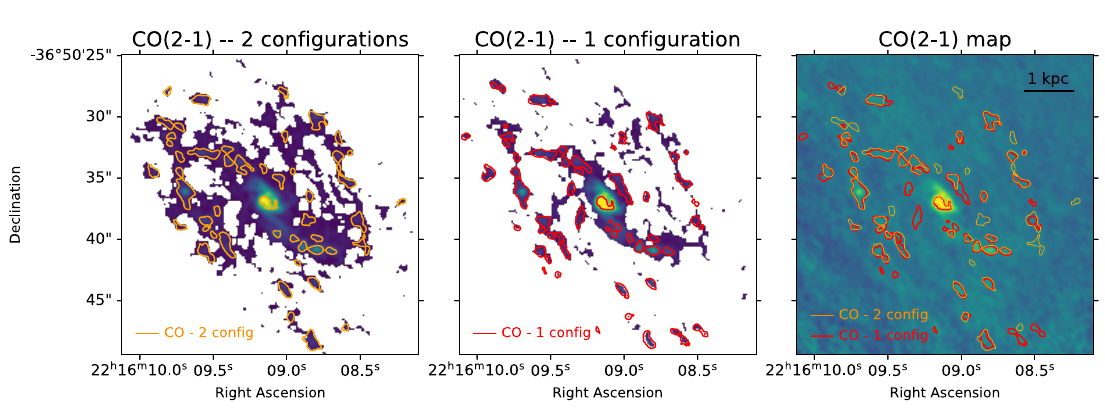}
   \caption{CO(2--1) integrated intensity maps of IC\,5179 at $\sim$100~pc scales, obtained with ALMA using two configurations (left panel) and only one extended configuration (middle panel). These maps illustrate, with orange and red contours, the identified regions from the dendrogram analysis detailed in Sect. \ref{sec:astrodedro}, applied to each map based on the same criteria. The map without clipping ({\it right panel}) displays the clumps identified.
   from observations using both one and two configurations. 
   } 
              \label{fig:ap1}
  \end{figure*}

To evaluate the influence of using one or two configurations on clump detection, we compared the identified regions (clumps) in observations of IC\,5179 using different configurations, as shown in Figure~\ref{fig:ap1}.

We find that most of the clumps identified in the two-configuration observations are also detected in the single-configuration observations. This result suggests that the identification of clumps in observations with different configurations does not significantly impact the study of the star formation process in molecular clumps.

In addition, we compared the integrated fluxes within the identified clumps using the different array setups. For IC~5179, the flux difference between the two 12 m configurations corresponds to a modest loss of $\sim$8.5\%. As a consistency check, we also compared a single 12m configuration with the combined 12m+7m data in NGC~7469, finding a flux difference of $\sim$23.2\%. These tests confirm that, to a large extent, the overall clump fluxes are preserved within typical uncertainties, and our statistical results remain unaffected.

\section{Comparison and effects of the selected CO-to-H$_{2}$ conversion factor}  \label{app:conversionfactor}

We used a fixed Galactic CO-to-H$_{2}$ conversion factor to estimate the molecular gas mass of our sample. However, this factor depends on the properties of the gas, including temperature, density, and metallicity. Its precise value remains a topic of debate and can vary significantly across different environments, particularly in extreme conditions such as those found in (U)LIRGs. 
Recent studies of star-forming regions within the PHANGS sample, at 90-150~pc scales, have revealed that the velocity dispersion of molecular gas significantly influences the value of $\alpha_{CO}$. For instance, \cite{Teng2023, Teng2024} reported a strong anti-correlation between $\alpha_{CO}$ and the local cloud-scale velocity dispersion. Higher velocity dispersion enhances the rate of collisional excitation of molecules, potentially leading to an overestimation of $\alpha_{CO}$. 

The fitted function, derived from a galaxy sample using velocity dispersion measurements at a $\sim$ 90~pc scale, is as follows:

\begin{equation}
  log_{10} \alpha_{co} = -0.59  (\pm 0.04) ~ log <\sigma_{v}>_{90pc} + ~0.90 (\pm 0.03)
  \label{eq:teng}
\end{equation}

By applying the $\sigma_{v}$-based $\alpha_{CO}$ prescription\footnote{Equation \eqref{eq:teng} follows the formulation from private communication and will be included in a forthcoming paper.} to our sample, we aim to verify that this $\alpha_{CO}$ prescription does not alter the results of this work. We use two approaches: 1) we use $\alpha_{CO}$ values derived from the equation, and 2) we assume a fixed value in the regime where the velocity dispersion is not covered by \cite{Teng2024}. 
This second approach is considered because the galaxies in our sample cover higher velocity dispersion values than the range explored in the PHANGS sample. The regime of high velocity dispersion values has not been explored in the $\sigma_{v}$-based $\alpha_{CO}$ prescription, which makes the behaviour of the factor with velocity dispersion uncertain. The PHANGS sample covers values of velocity dispersion up to $\sim$35~km/s, whereas the galaxies in our sample exceed this range of values.

 \begin{figure}[htbp!]
   \centering
   \includegraphics[width=8.3cm]{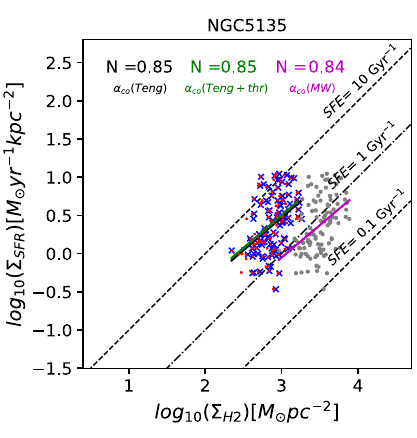}
   \caption{SFR surface density as a function of the molecular gas surface density at $\sim$100 pc in NGC~5135 using three different prescriptions in the conversion factor: $\sigma_{v}$-based $\alpha_{CO}$ prescription (black fit), the same prescription with a threshold in the region of velocity dispersion unexplored (green fit) and the Galactic conversion factor (magenta fit). 
   } 
              \label{fig:ap2}
  \end{figure}

As an example, Fig. \ref{fig:ap2} shows the KS plot for the galaxy NGC5135. We obtain similar results with the only difference of the lower values in the molecular gas surface density. For the case of dual galaxies, we still obtain the dual behaviour and the single power law for the galaxies using the clumps methodology. The only difference is in the range of values of the x-axis where the surface density of molecular gas is lower, with a shift of $\sim$ 0.55 dex. This implies that the depletion time changes by a factor of approximately 3.7. However, we can conclude that the different prescriptions for the conversion factor do not affect the main results of this work.

\section{Additional figures}

\subsection{Distribution of the clump identification and beam-sized region methods in the dual LIRG NGC 7130}\label{app:dual_2methods}
In this Appendix, we show as an example the spatial distribution of the regions and clumps in NGC\,7130, which exhibits dual behaviour. 

\begin{figure*}[t]
   \centering
    \includegraphics[width=.8\linewidth]{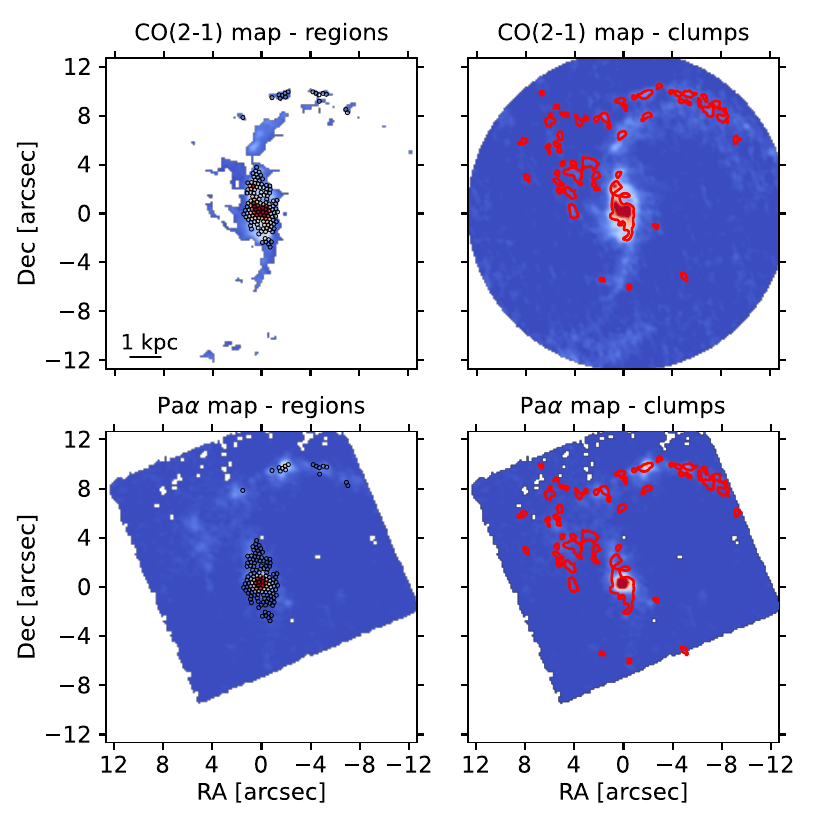}
 \caption{ALMA CO(2--1) integrated intensity (moment 0) map and HST Pa$\alpha$ image of the galaxy NGC~7130. Left: Location of the
beam-sized regions (in black) on the CO(2--1) (top) and Pa$\alpha$ (bottom) maps. Right: Location of physical structures (in red) found using Astrodendro (clumps) on the CO(2--1) (top) map, also projected over the Pa$\alpha$ (bottom) map. The regions and clumps are both detected with ALMA and HST}.
    \label{fig:regions_dual_2methods}
\end{figure*}

\subsection{Spatial distribution of the dual behaviour in the KS diagram for the LIRG NGC 3110}\label{app:ksplots}
In this Appendix, we show as an example the spatial locations of the regions in  NGC\,3110, which exhibits dual behaviour. 

\begin{figure}[ht!]
   \centering
    \includegraphics[width=.95\linewidth]{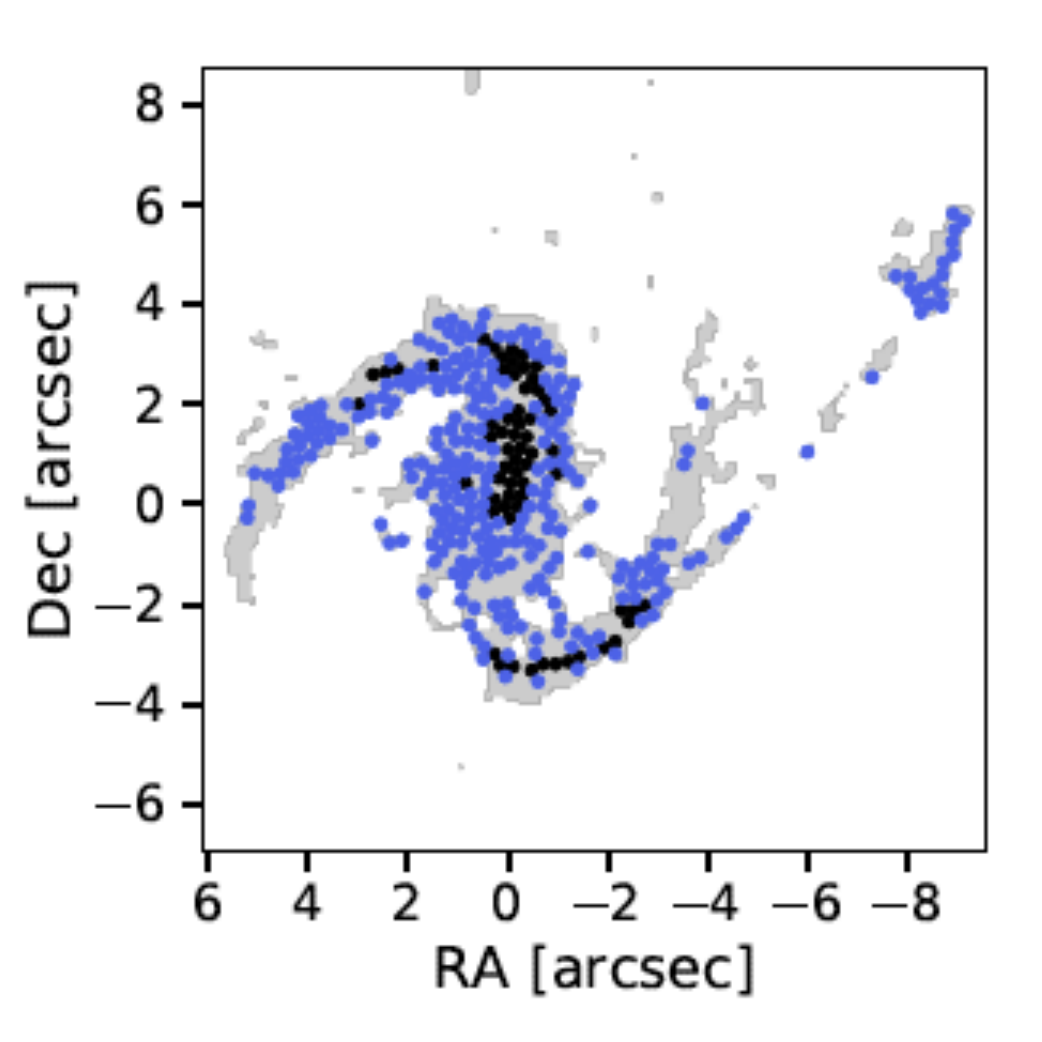}
 \caption{Location of the regions on the CO(2–1) map (grey) for NGC~3110. The black and blue points indicate regions corresponding to the two branches of the KS relation shown in Fig. \ref{fig:ksplot}: blue points correspond to higher $\Sigma_{H_2}$ and $\Sigma_{SFR}$, while black points correspond to lower $\Sigma_{H_2}$ and $\Sigma_{SFR}$}.
    \label{fig:regions_dual}
\end{figure}

\subsection{Depletion time, star formation efficiency, self-gravity and velocity dispersion in individual galaxies}\label{app:parameters_plots}
In this Appendix, we show, as an example, the scatter plots of the different parameters for the galaxy NGC 7469, which complement Figure \ref{fig:other_parameters_individual_galaxies}.

\begin{figure*}[h]
   \centering
    \makebox[\textwidth][c]{%
\includegraphics[width=1.05\textwidth]{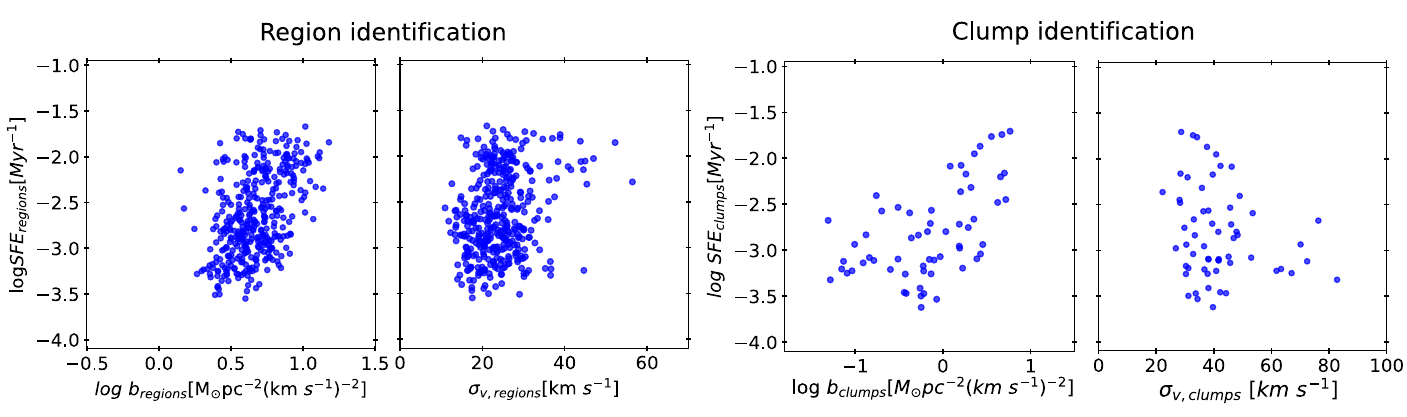}}
 \caption{Scatter plots for NGC 7469 showing the star formation efficiency (SFE) as a function of the boundedness parameter ($b$), and of the velocity dispersion ($\sigma_{v}$). From left to right, the panels show: SFE vs. 
$b$ and SFE vs. $\sigma_{v}$ for beam-sized regions, followed by SFE vs. $b$ and SFE vs. $\sigma_{v}$ for clump identification.}
    \label{fig:parameters_plots}
\end{figure*}

\section{Star formation properties as a function of merger stage using beam-sized regions}\label{app:regions_measurements}

\begin{figure}[ht!]
   \centering
    \includegraphics[width=.825\linewidth]{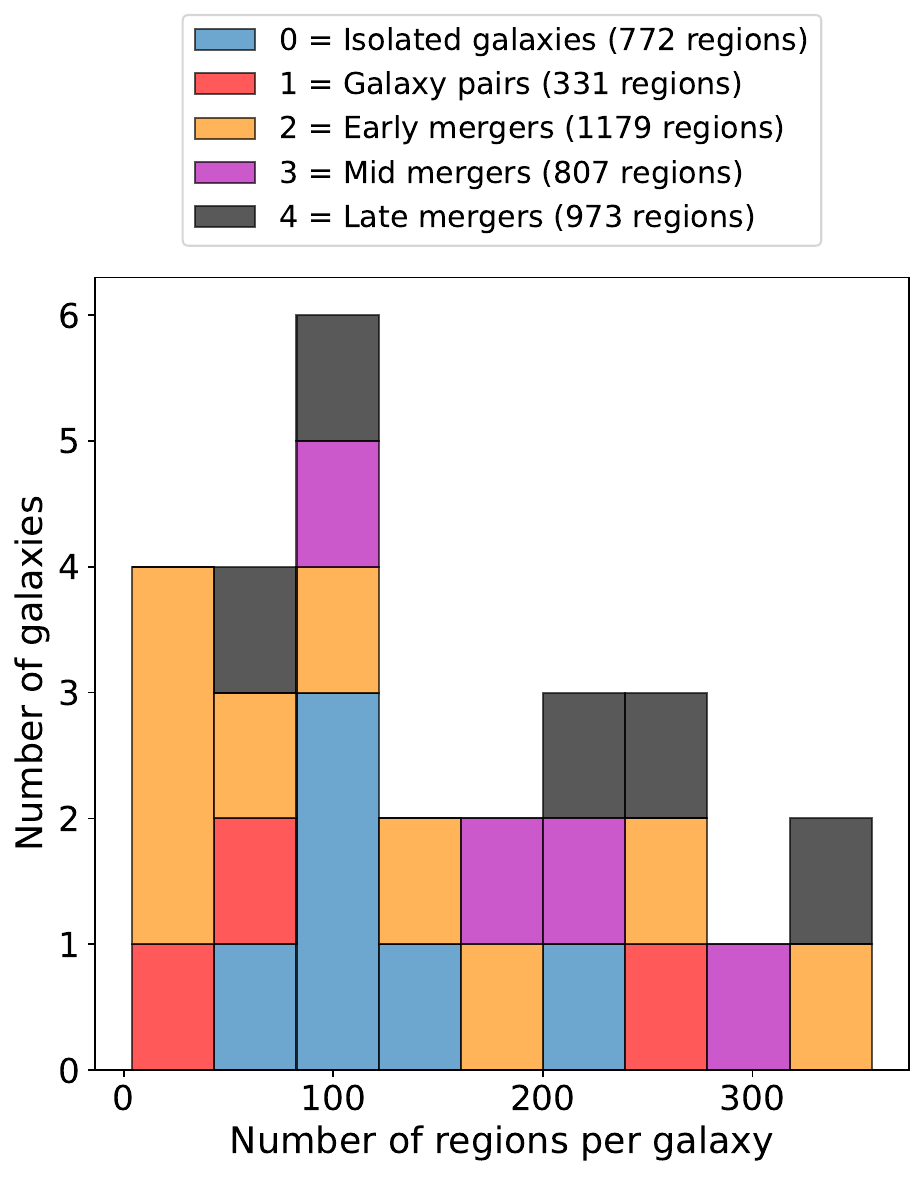}
 \caption{Distribution of the number of regions per galaxy across the different merger stages in the galaxy sample. The early-, mid-, and late-stage mergers cover the entire range of the stacked histogram, while isolated galaxies show lower number of regions per galaxy.
 }
    \label{fig:histogramregions}
\end{figure}

\begin{figure*}[ht]
    \centering
    \includegraphics[width=.79\linewidth]{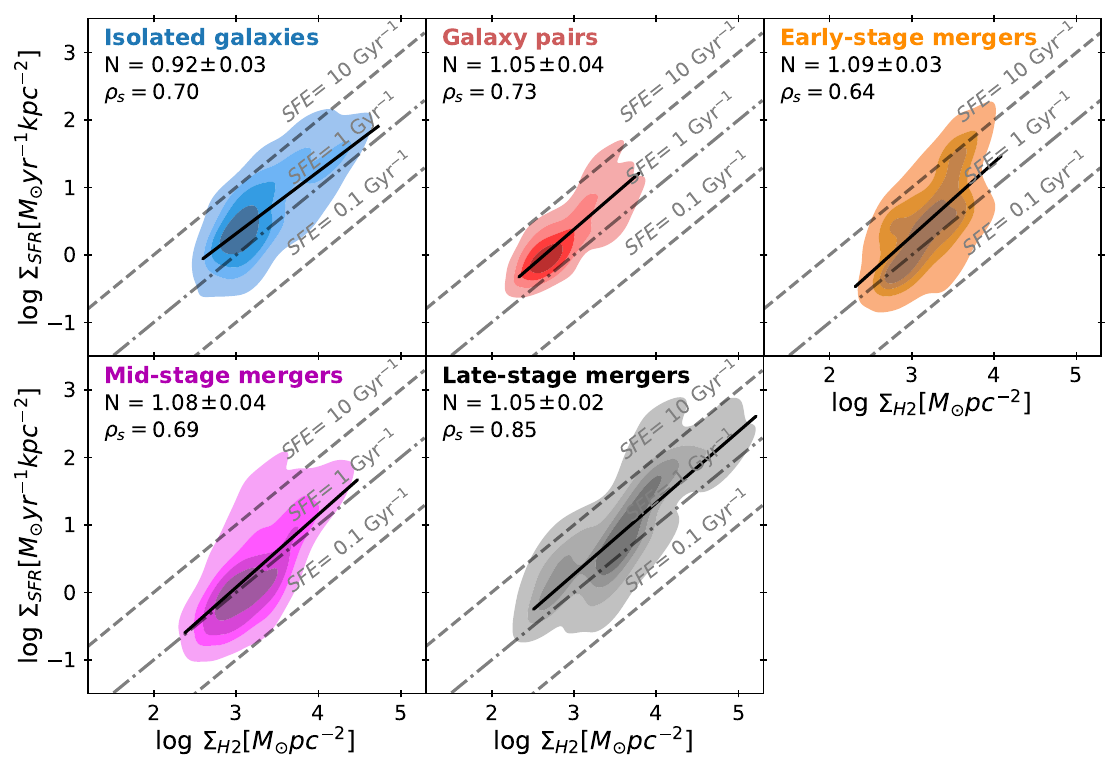}
    
    \caption{Representation of the KS diagram across the merger sequence  based on beam-sized region identification. From left to right and top to bottom, the merger stages are: isolated galaxies (blue), galaxy pairs (red), early-stage mergers (orange), mid-stage mergers (magenta) and late-stage mergers (black). The black solid line represents the best fit for each dataset. The Spearman’s rank correlation coefficients ($\rho_{s}$) and the power-law indices (N) of the derived best-fit KS relations are indicated. The bottom right panel shows the mean values of log$_{10}\Sigma_{SFR}$ and log$_{10}\Sigma_{H2}$, with error bars representing the mean absolute deviation for each merger stage. The grey dashed lines mark constant star formation efficiencies (SFE = $\Sigma_{SFR}$/$\Sigma_{H2}$). 
    }
    \label{fig:ksplot_merger_sequence_regions}
\end{figure*}

\begin{figure*}[ht]
    \centering
\includegraphics[width=.79\linewidth]{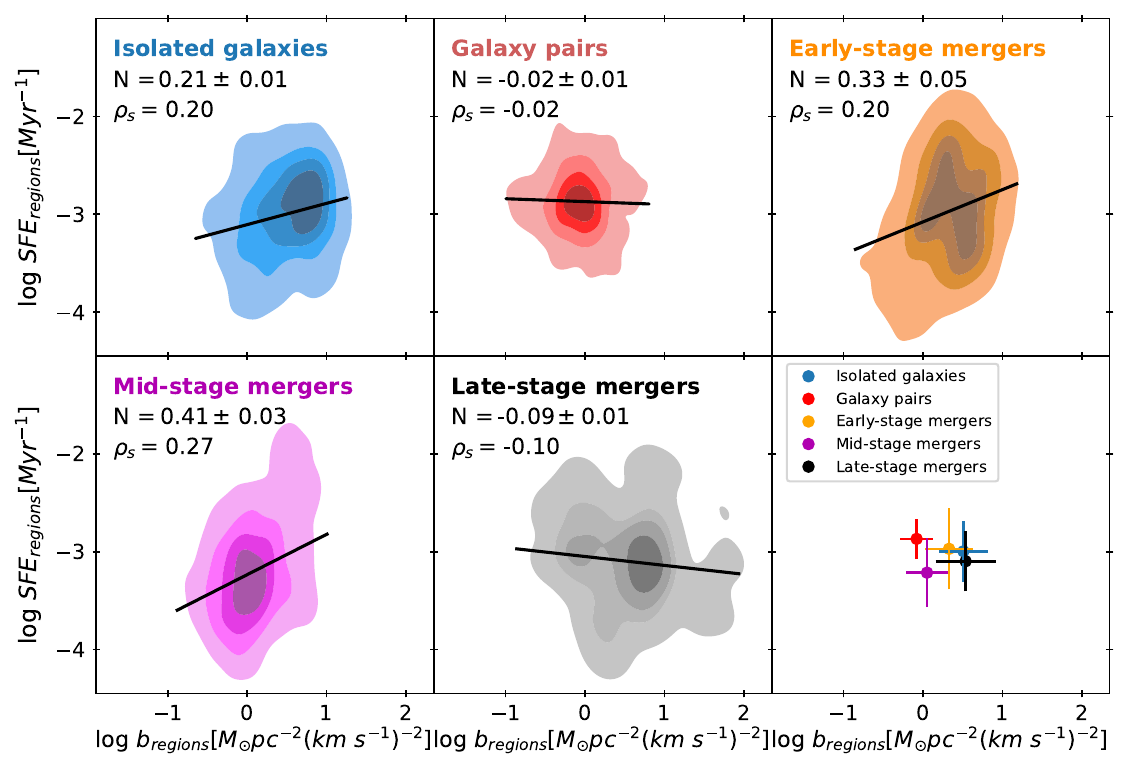}
    \caption{SF efficiency, SFE as a function of the boundedness of the regions (parameter $b_{regions}$) across the merger sequence. From left to right and top to bottom, the merger stages are: isolated galaxies (blue), galaxy pairs (red), early-stage mergers (orange), mid-stage mergers (magenta) and late-stage mergers (black). The Spearman’s rank correlation coefficients ($\rho_{s}$) and the power-law indices (N) of the best-fit relations are indicated. The bottom right panel shows the mean values of log$_{10}SFE_{regions}$ and log$_{10}b_{regions}$, with error bars representing the mean absolute deviation for each merger stage. 
    }
    \label{fig:b_merger_sequence_regions}
\end{figure*}

\begin{figure*}[ht]
    \centering
\includegraphics[width=.79\linewidth]{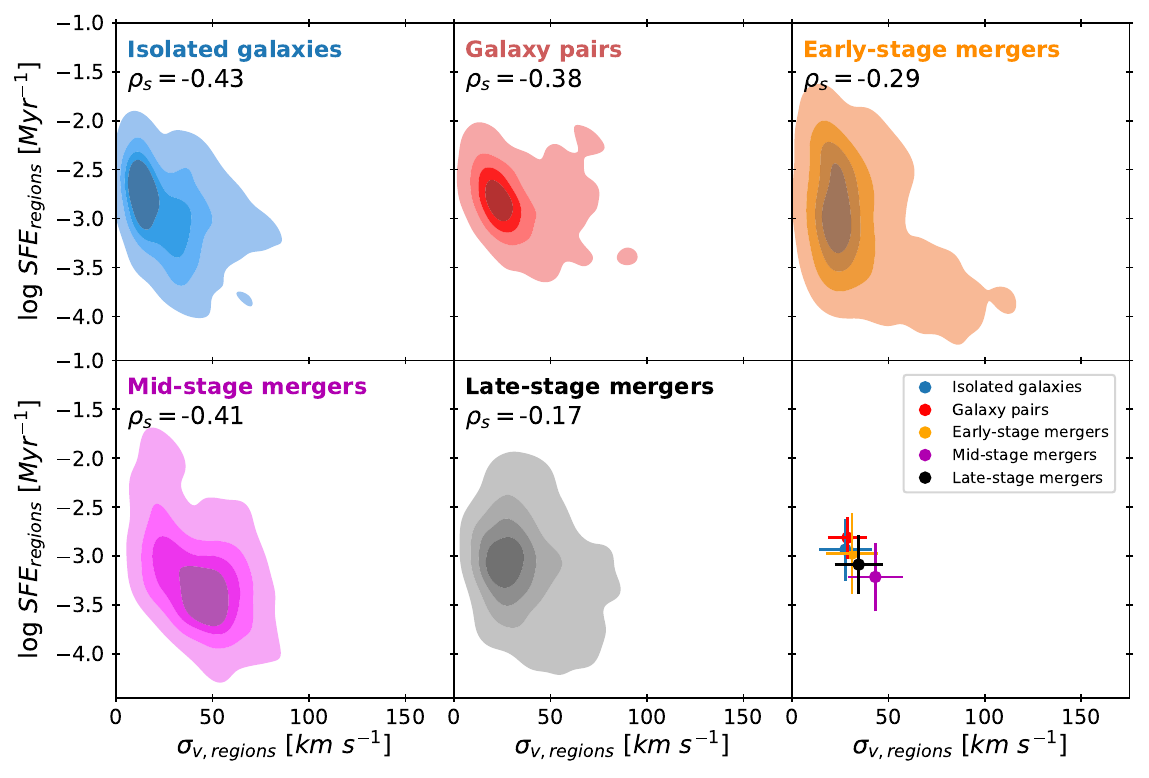}
    \caption{SF efficiency as a function of the velocity dispersion of the regions ($\sigma_{v,regions}$) across the merger sequence. From left to right and top to bottom, the merger stages are: isolated galaxies (blue), galaxy pairs (red), early-stage mergers (orange), mid-stage mergers (magenta) and late-stage mergers (black). The Spearman’s rank correlation coefficients ($\rho_{s}$) are indicated. The bottom right panel shows the mean values of $\sigma_{v,regions}$ and $log_{10}SFE_{regions}$, with error bars representing the mean absolute deviation for each merger stage. 
    }
    \label{fig:turbulence_merger_sequence}
\end{figure*}

We present the study of star‑formation properties in the sample using the beam‑sized region methodology. Figure \ref{fig:histogramregions} shows the distribution of regions across merger stages; Figure \ref{fig:ksplot_merger_sequence_regions} presents the Kennicutt-Schmidt law; Figure \ref{fig:b_merger_sequence_regions} displays the star formation efficiency as a function of the self‑gravity of the regions; and Figure \ref{fig:turbulence_merger_sequence} shows the star formation efficiency as a function of the velocity dispersion of the regions.

When exploring the KS relation using beam-sized regions (see Figure \ref{fig:ksplot_merger_sequence_regions}), we find that the slope values remain close to one in all cases and do not vary significantly throughout the merger sequence. This suggests that, when considering the gas and star-forming emission of galaxies without identifying physical structures such as clumps or clouds, the relationship between the surface densities of gas and SFR does not evolve with the merger stage. 
The uniform slope observed using the beam-sized regions method likely reflects the influence of the larger gas disc across the full extent of the galaxy, as these unresolved lines-of-sight include both diffuse emission and denser regions, smoothing out the variations detected when focusing on clumps.

When exploring the star formation efficiency using beam-sized regions (Figure \ref{fig:b_merger_sequence_regions}), we find that the trends are generally more uniform across merger stages. This discrepancy with the clump-based method arises because the region-based approach examines gas emission and star formation rates across the entire galaxy, rather than focusing on physically coherent structures. The region selection provides more information about the dynamic environment of molecular clouds, offering insights into inter-cloud turbulence (see Sect. \ref{velocityd}), whereas the clump approach targets the properties of structures where stars are currently forming or will form in the future.
This detailed comparison emphasises how different analysis methods probe different physical scales, with clumps highlighting local star formation efficiency and beam-sized regions capturing large-scale disc effects.

We present the SFE versus velocity dispersion using beam-sized regions ($\sigma_{v,beam}$) across the merger sequence (Figure \ref{fig:turbulence_merger_sequence}). The trends are generally smoother and less scattered than for clumps.
 Comparing these approaches illustrates that the physical scale of the measurement strongly influences the observed correlation between SFE and velocity dispersion. The complementary nature of these two methods provides a more complete picture of how gas dynamics regulate star formation during galaxy mergers.

\section{Comparison between region-based and clump-based measurements}\label{app:comparisons}

\begin{figure*}[ht]
    \centering

    \includegraphics[width=.79\linewidth]{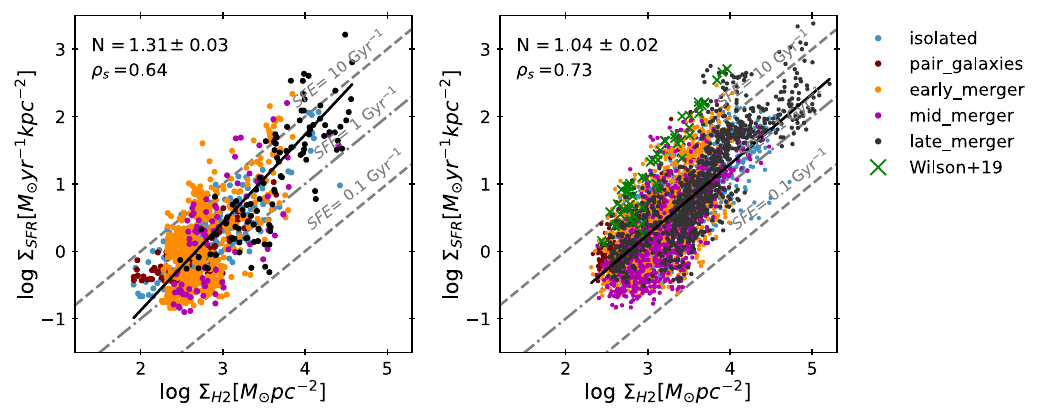}
    
    \caption{KS relations for clumps (left) and regions (right) in our sample. Clumps show a superlinear trend, whereas regions yield a nearly linear relation. Data from the U/LIRG study of \cite{wilson2019} are shown in green. The different colours correspond to the merger stages of the galaxies: isolated (blue), pairs (red), early-stage mergers (orange), mid-stage mergers (magenta), and late-stage mergers (black). The Spearman’s rank correlation coefficients ($\rho_{s}$) and the power-law indices (N) of the best-fit relations, derived from our sample, are indicated.}
    \label{fig:total_regions_clumps}
\end{figure*}

Figure \ref{fig:total_regions_clumps} shows the KS relation for all the clumps and regions in our LIRGs sample. 

Our clump-based analysis (left panel) yields a superlinear slope (N = 1.31 $\pm$ 0.03, $\rho_{s}$ = 0.64), whereas the region-based analysis (right panel) results in a slope close to linear (N = 1.04 $\pm$ 0.02, $\rho_{s}$ = 0.73).  
We quantify the scatter around the best-fit K–S relation using the standard deviation of the orthogonal distances between the data points and the ODR fit, finding 0.21 dex for the clump-based measurements and 0.28 dex for the region-based ones. This difference indicates that clumps follow a more tightly correlated KS relation, with a scatter approximately 25\% lower than that of regions.  
The two methods reveal variations in both the slope and the correlation coefficient, providing complementary perspectives on the star formation relation: one emphasising the statistics of gas and SF properties at the smallest scales, and the other targeting physically coherent gas clumps and structures. These differences may reflect variations in the dominant physical processes at different spatial scales, including the potential role of disc pressure in regulating star formation in addition to cloud self-gravity. 

We further included the \cite{wilson2019} data (right panel), a sample of five U/LIRGs at scales between 345 -- 650 pc, which fall within the range spanned by our galaxy sample in the plot. Our beam-based regions results are consistent with previous works, where the slope of the KS relation generally converges toward  N $\approx$ 1 \citep[e.g.][]{Kennicutt2021}. \cite{wilson2019} reported a steeper slope of N = 1.73, adopting a constant CO-to-H$_{2}$ conversion factor of $\sim$0.8 M$_{\odot}$(K km s$^{-1}$ pc$^{2}$)$^{-1}$, about five times lower than the standard value for self-gravitating molecular clouds \citep{downes1998}. 
Using the same $\alpha_{\mathrm{CO}}$ for our galaxies yields an identical slope, but shifted by $\sim$0.72 dex along the x-axis, and with depletion times shorter by a factor of $\sim$5. 
If such a low $\alpha_{\mathrm{CO}}$ is only adopted for the nuclear regions of our galaxies, the contrast with the outer regions becomes more pronounced, thereby reinforcing the dual behaviour. 

Interestingly, we find that two galaxies in common with \cite{wilson2019} exhibit superlinear slopes in our study of individual galaxies, 
consistent with the results obtained by \cite{wilson2019} when analysing their full sample of five galaxies. This suggests that some of the remaining galaxies in their sample could also exhibit superlinear slopes, which would explain why the combined slope of their sample reaches $\sim$1.73. In contrast, when all regions from all galaxies in our sample are considered together, at a resolution comparable to that of molecular clouds, the KS relation tends to be linear, highlighting the crucial role of both spatial resolution and sample size in determining the observed slope. At spatial scales of $\sim$1.5 kpc in 80 nearby galaxies, \citet{sun2023} also report slopes in the range 0.9--1.2, consistent with the values obtained in this work.

\begin{figure*}[ht]
    \centering



  \includegraphics[width=.879\linewidth]{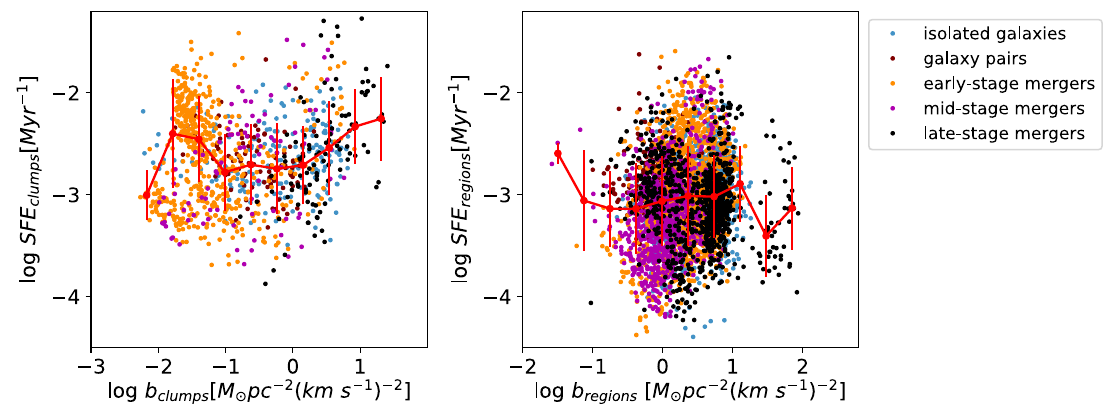}
    
    \caption{SF efficiency (SFE) as a function of the self-gravity parameter (\textit{b}), shown for clumps (left) and regions (right) in our sample. Clumps display a weak trend with increasing self-gravity, whereas regions show a flat relation. When considering the full sample, the scatter is large, particularly for regions. Colours indicate the merger stage of the galaxies: isolated (blue), pairs (red), early-stage mergers (orange), mid-stage mergers (magenta), and late-stage mergers (black). 
    }
    \label{fig:tdep_b_total_regions_clumps}
\end{figure*}

\begin{figure*}[ht]
    \centering

   \includegraphics[width=.871\linewidth]{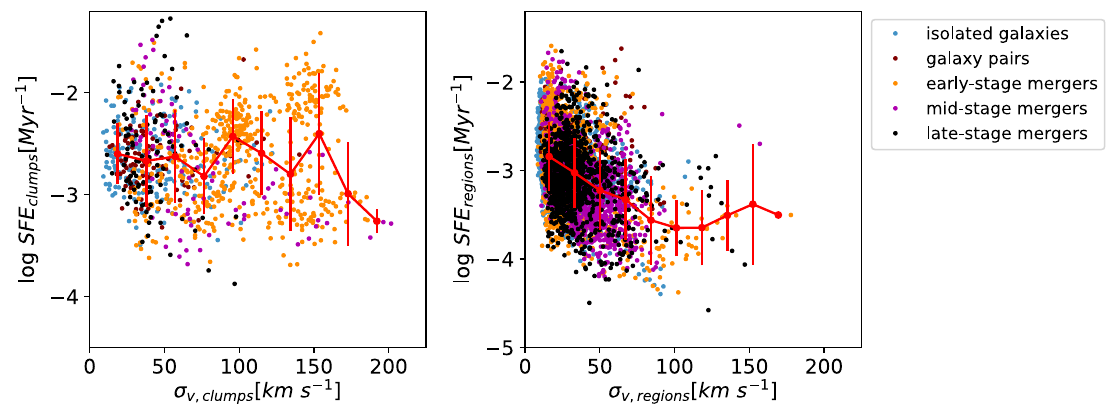}
    
    \caption{SF efficiency as a function of the velocity dispersion, shown for clumps (left) and regions (right) in our sample. Clumps exhibit a flat relation, while regions show a weak anticorrelation with increasing velocity dispersion. Considering the full sample, the scatter is large. Colours indicate the merger stage of the galaxies: isolated (blue), pairs (red), early-stage mergers (orange), mid-stage mergers (magenta), and late-stage mergers (black). 
    }
    \label{fig:sfe_sigma_total_regions_clumps}
\end{figure*}

In addition, when analysing the SFE, self-gravity, and velocity dispersion across the full sample (Figures \ref{fig:tdep_b_total_regions_clumps} and \ref{fig:sfe_sigma_total_regions_clumps}), we find that the correlations remain weak and highly scattered. Clumps display a weak correlation at high $b$ values, while regions show no significant trends. Clumps show an approximately flat SFE-$\sigma_{v}$ relation, whereas regions display a slight anticorrelation with increasing velocity dispersion of the gas. The scatter is large when considering all data points.

\end{appendix}

\end{document}